\documentclass{aa}  

\usepackage{graphicx}
\usepackage{txfonts}
\usepackage{amsmath}
\usepackage{amssymb}
\usepackage{natbib}

\graphicspath{{./figures/}}
\bibpunct{(}{)}{;}{a}{}{,}
\newlength{\minuslength}
\begin{document}

   \title{Fractionation in young cores: Direct determinations of nitrogen and carbon fractionation in HCN}


   \author{S. S. Jensen\inst{1}\thanks{\email{sigurdsj@mpe.mpg.de}}
                \and S. Spezzano \inst{1}
          \and P. Caselli \inst{1}
          \and O. Sipil{\"a}\inst{1}
          \and E. Redaelli\inst{1}
          \and K. Giers \inst{1}
          \and J. Ferrer Asensio\inst{1} 
        }

   \institute{Max-Planck-Institut f{\"u}r extraterrestrische Physik, Giessenbachstrasse 1, D-85748 Garching, Germany}

   \date{Draft date: \today}

  \abstract
  {Nitrogen fractionation is a powerful tracer of the chemical evolution during star and planet formation. It requires robust determinations of the nitrogen fractionation across different evolutionary stages.}
  {We aim to determine the ${}^{14}$N/${}^{15}$N and ${}^{12}$C/${}^{13}$C ratios for HCN in six starless and prestellar cores and to compare the results between the direct method using radiative transfer modeling and the indirect double isotope method, assuming a fixed ${}^{12}$C/${}^{13}$C ratio.}
  {We present IRAM observations of the HCN $1$-$0$, HCN 3-2, HC$^{15}$N 1-0 and H$^{13}$CN $1$-$0$ transitions toward six embedded cores. The ${}^{14}$N/${}^{15}$N ratio was derived using both the indirect double isotope method and directly through non-local thermodynamic equilibrium (NLTE) 1D radiative transfer modeling of the HCN emission. The latter also provides the ${}^{12}$C/${}^{13}$C ratio, which we compared to the local interstellar value.} 
  {The derived ${}^{14}$N/${}^{15}$N ratios using the indirect method are generally in the range of 300-550. This result could suggest an evolutionary trend in the nitrogen fractionation of HCN between starless cores and later stages of the star formation process. However, the direct method reveals lower fractionation ratios of around $\sim$250, mainly resulting from a lower ${}^{12}$C/${}^{13}$C ratio in the range $\sim$20--40, as compared to the local interstellar medium value of 68.}
  {This study reveals a significant difference between the nitrogen fractionation ratio in HCN derived using direct and indirect methods. This can influence the interpretation of the chemical evolution and reveal the pitfalls of the indirect double isotope method for fractionation studies. However, the direct method is challenging, as it requires well-constrained source models to produce accurate results. 
  No trend in the nitrogen fractionation of HCN between earlier and later stages of the star formation process is evident when the results of the direct method are considered.}

   \keywords{astrochemistry ---
                stars: formation ---
                ISM: abundances ---
                submillimeter: stars ---
                ISM: individual objects: CB23, TMC2, L1495, L1495AN, L1512, L1517B
               }

   \maketitle
%
\section{Introduction}
Isotopic fractionation is an important tool to study interstellar chemistry and trace the evolution of material from molecular clouds to planetary systems \citep[e.g.,][]{2012A&ARv..20...56C, 2014Sci...345.1590C, 2018A&A...609A.129C, 2021A&A...650A.172J}. 
In particular, the nitrogen isotopic ratio $^{14}$N/$^{15}$N can trace the evolution of primordial Solar System material up to the present day \citep[e.g.,][]{2014A&A...572A..24W, 2018MNRAS.474.3720W}. 

The $^{14}$N/$^{15}$N ratio in the primitive solar nebula has been derived from measurements of the solar wind, with a result of 440 \citep{2011Sci...332.1533M}. While a consistent ratio was found in the atmosphere of Jupiter \citep[$^{14}$N/$^{15}$N=450, ][]{2004Icar..172...50F}, an enrichment of $^{15}$N is observed towards the terrestrial atmosphere \citep[$^{14}$N/$^{15}$N=272, ][]{1950PhRv...77..789N} and comets  \citep[$^{14}$N/$^{15}$N=139, ][]{2014ApJ...782L..16S}.
The Solar and Jovian nitrogen isotope ratios are believed to represent the composition of the protosolar nebula, whereas the origin of the $^{15}$N-enrichment in terrestrial planets atmospheres, comets, and some meteorites is not yet fully understood. 
In the interstellar medium (ISM), the  $^{14}$N/$^{15}$N ratio varies among different sources as well as with different molecular tracers. The $^{14}$N/$^{15}$N ratio in N$_2$H$^+$ towards starless cores has been observed to be as high as 1000 \citep{2013A&A...555A.109B, 2018A&A...617A...7R}, while lower values of $\sim$420 are observed toward protostars \citep{2020A&A...644A..29R}. The $^{14}$N/$^{15}$N ratio in ammonia ranges between 210 and 800 in starless and protostellar cores \citep{2009A&A...498L...9G, 2023A&A...674L...8R}. The $^{14}$N/$^{15}$N ratio in nitriles (CN, HCN, and HNC) observed towards low-mass starless cores ranges between 140 and 360 \citep[][ and references therein]{2020A&A...643A..76H}. 

\begin{figure*}[ht]
\resizebox{\hsize}{!}
        {\includegraphics{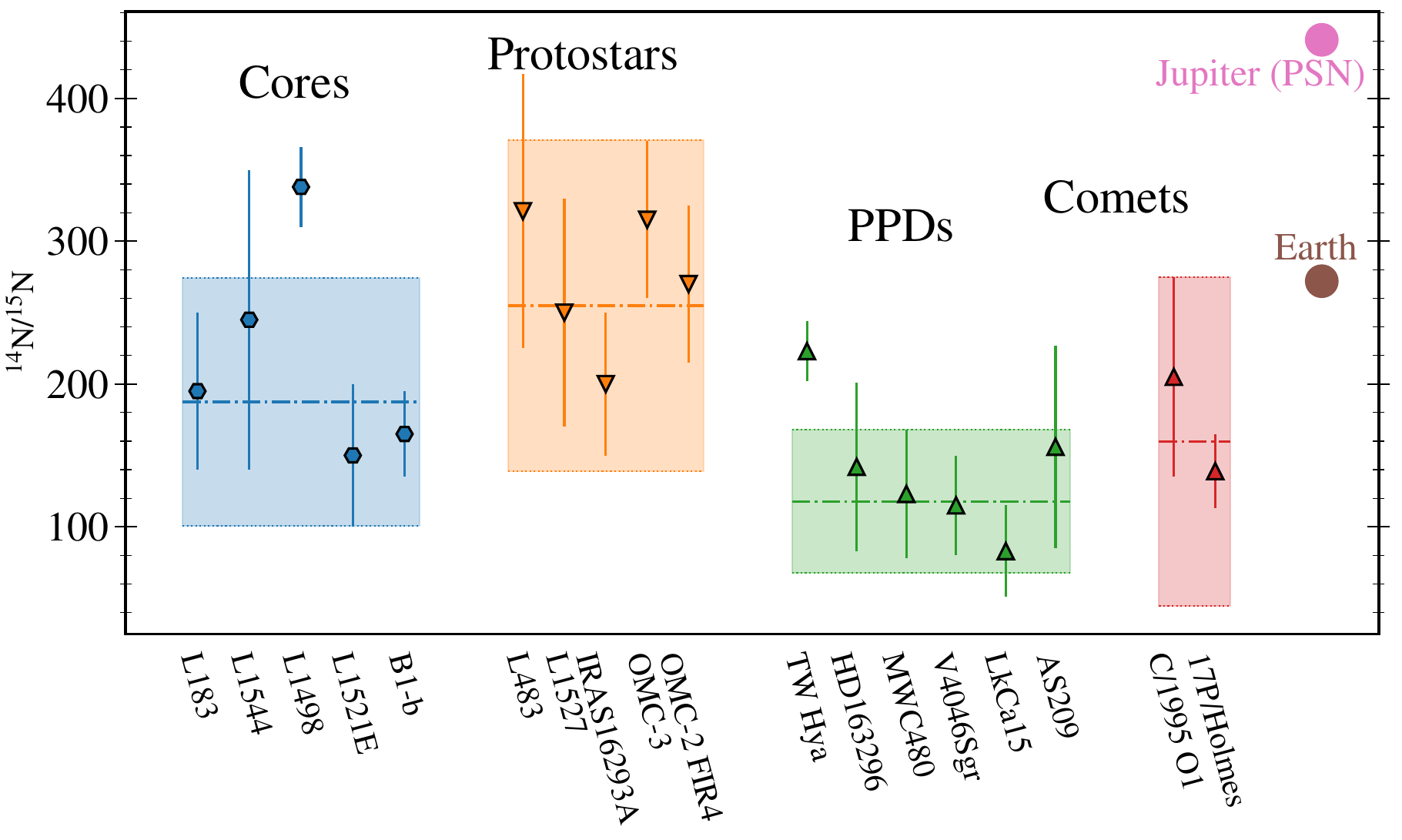}}
  \caption{Overview of reported $^{14}$N/$^{15}$N ratios in HCN at various evolutionary stages from cores to disks. The majority of the measurements for HCN are derived using the indirect double isotope method; hence, a fixed $^{12}$C/$^{13}$C ratio of 68 is assumed. Protoplanetary disks are abbreviated to PPDs in the figure. References are provided in Appendix \ref{app:references}.}
     \label{fig:Nfrac}
\end{figure*}

An overview of the nitrogen fractionation in HCN is presented in Fig. \ref{fig:Nfrac}. Observations cover all evolutionary stages from the starless core stages through the protoplanetary disk stage and, finally, the cometary values from the Solar System. 
Nonetheless, the current observational sample has proven insufficient in determining what physical and chemical processes regulate the degree of fractionation in nitrogen-bearing species and how the fractionation evolves throughout star and planet formation. Both environmental effects (e.g., differences in irradiation and temperature) and nucleosynthetic effects are possible drivers of the fractionation \citep[see, e.g.,][ and references therein]{2013A&A...557A..65H, 2018ApJ...857..105F, 2018MNRAS.478.3693C}. \citet{2022A&A...664L...2S} mapped the column densities of nitriles in the pre-stellar core L1544 and displayed local variations in the $^{14}$N/$^{15}$N ratio of HCN. These variations are consistent with a local gradient in the irradiation of the core, which could indicate that isotope-selective photodissociation plays a key role in the fractionation of nitrogen. Deciphering the underlying processes will provide a significant input to the study of chemical evolution during star- and planet formation as well as volatile delivery in the Solar System.

Previous measurements of the $^{14}$N/$^{15}$N ratio in starless and pre-stellar cores have primarily used the indirect double isotope method. This method combines observations of H$^{13}$CN and HC$^{15}$N and assumes a fixed $^{12}$C/$^{13}$C ratio to derive the abundance of the main isotopolog H$^{12}$C$^{14}$N and the $^{14}$N/$^{15}$N ratio. The advantage of this method is that observations of the optically thick HCN are avoided. However, the assumption of a fixed $^{12}$C/$^{13}$C ratio is uncertain. Recent chemical models studying the evolution of the $^{12}$C/$^{13}$C ratio in young cores have suggested that the ratio may vary significantly over time and may also depend on the local cosmic-ray ionization rate \citep{2020A&A...640A..51C, 2020MNRAS.498.4663L}.

As an alternative to the indirect method, the HCN abundance can be directly determined through non-Local Thermodynamic Equilibrium (NLTE) radiative transfer modeling. \citet{2013A&A...560A...3D} determined the the $^{12}$C/$^{13}$C and $^{14}$N/$^{15}$N ratios in Barnard B1b region by combining observations of multiple HCN, H$^{13}$CN, and HC$^{15}$N transitions with one-dimensional (1D) NLTE radiative transfer modeling. They found a $^{12}$C/$^{13}$C ratio of $\sim$30 and a $^{14}$N/$^{15}$N ratio of $\sim$165 in HCN.
Similarly, \citet{2018A&A...615A..52M} performed a direct determination of the $^{14}$N/$^{15}$N ratio toward the starless core L1498, by combining observations of the 1--0 and 3--2 transition of HCN with observations of the 1--0 transitions of  H$^{13}$CN and HC$^{15}$N and performing NLTE radiative transfer modeling. That study found that the $^{14}$N/$^{15}$N ratio is consistent with the elemental value in L1495, suggesting that no fractionation of nitrogen had occurred, while a $^{12}$C/$^{13}$C ratio of 45$\pm3$ was derived, suggesting instead that carbon is fractionated in the core and not consistent with the elemental ISM value of 68 \citep{2005ApJ...634.1126M}. 
A third method to derive the $^{14}$N/$^{15}$N and/or $^{12}$C/$^{13}$C ratio of HCN is to observe the double substituted H$^{13}$C$^{15}$N, along with HC$^{15}$N and/or H$^{13}$CN. Since these isotopologs are generally optically thin, circumventing issues related to optical depth of the main HCN isotopolog. H$^{13}$C$^{15}$N was recently observed in the dense core surrounding the class 0 source L483. In \citet{2019A&A...625A.147A},  $^{14}$N/$^{15}$N = 321$\pm$96 and $^{12}$C/$^{13}$C = 34$\pm$10 was found in HCN. This method has the advantage that it does not require radiative transfer modeling of the core as the direct method, but instead requires enough sensitivity and spectral resolution to observe the weaker H$^{13}$C$^{15}$N isotopolog at sufficiently high signal-to-noise ratio (S/N).

In this work, we present both indirect and direct determinations of the $^{14}$N/$^{15}$N and $^{12}$C/$^{13}$C ratios for five starless cores and one pre-stellar core. The classification of the evolutionary stage of the cores is taken from \citet{2005ApJ...619..379C}, where the more evolved cores with higher central densities, higher N$_2$D$^{+}$/N$_2$H$^{+}$ ratio, and higher CO depletion are characterized as pre-stellar. The primary aim is to increase the sample of reported $^{14}$N/$^{15}$N and $^{12}$C/$^{13}$C ratios in starless and pre-stellar cores to determine how much variation is present at this evolutionary stage. Furthermore, we also compare the direct and indirect methods and discuss the reliability of the indirect method when determining the $^{14}$N/$^{15}$N ratio of HCN.

\section{Observations} \label{sec:2}
Six cores were observed with the IRAM 30m single-dish (Pico Veleta, Spain) in 2022. The ground-state 1-0 transition of HCN, H${}^{13}$CN, and HC${}^{15}$N were observed with the EMIR E090 receiver, using the Fourier Transform Spectrometer (FTS) with a resolution of 50 kHz. Observations of the 3-2 transition of HCN were performed with the E230 receiver using both the FTS (50~kHz resolution) and the VESPA (20~kHz resolution) backends simultaneously. However, issues with the VESPA backend resulted in low S/N and, consequently, this work relies on the FTS data.
For each source, a single pointing toward the dust continuum peak was observed. The source sample was selected based on bright HCN 1--0 detections in a previous dataset \citep{2018A&A...609A.129C} and \emph{Herschel}/SPIRE coverage to determine the density profiles as described in the next section.
An overview of the observations is provided in the log presented in Table 1.
Frequency switching was performed with a frequency throw of $\pm±3.9$~MHz. Telescope pointing was checked every 1.5-2 hours and focus every 3-4 hours depending on the weather conditions.
The data were calibrated and continuum-subtracted using the {\sc GILDAS} software \footnote{\url{https://www.iram.fr/IRAMFR/GILDAS}}.

\begin{table}
\centering\caption{Observation summary. All sources are located in Taurus ($d \sim 140$~pc). Distances to the individual sources within Taurus may vary up to $\sim$10\% \citep{2018ApJ...859...33G}, but this does not impact the relative ratios presented here. All sources are starless cores, except TMC2, which is classified as pre-stellar.}             
\label{table:observations}
\centering          
\begin{tabular}{c c c c c c c}  
\hline \hline       
            \noalign{\smallskip}
Source & R.A. (J2000) & DEC (J2000) & v$_\mathrm{LSR} $(km/s)\\  
            \noalign{\smallskip}
\hline                   
            \noalign{\smallskip} 
            CB23 & 04:43:27.7 & 29:39:11.0 &  6.0  \\
            TMC2  & 04:32:48.7 & 24:25:12.0 &  6.2 \\
            L1495  & 04:14:08.2 & 28:08:16.0  & 6.8 \\
            L1495AN  & 04:18:31.8 & 28:27:30.0  & 7.3 \\
            L1512  & 05:04:09.7 & 32:43:09.0 &  7.1 \\
            L1517B  & 04:55:18.8 & 30:38:04.0 & 5.8 \\

            \noalign{\smallskip} 
\hline                                    
\end{tabular}
\end{table}

\section{Results} \label{sec:3}
\begin{figure}[ht]
\resizebox{\hsize}{!}
        {\includegraphics{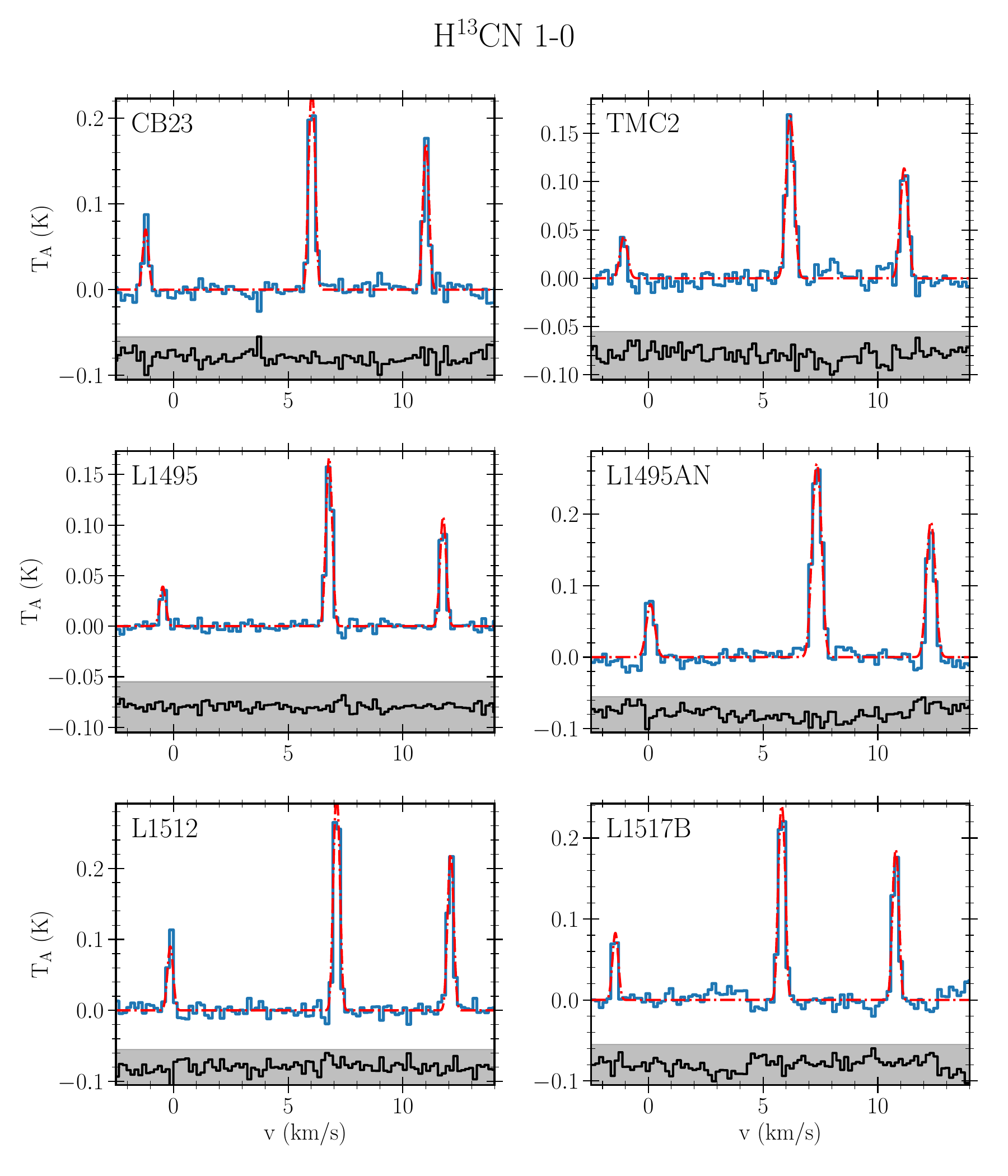}}
  \caption{Observed spectra for the 1--0 transition of H${}^{13}$CN for the six cores. The best fit to the hyperfine structure is shown in red and residuals between the fit and the observed spectra are shown in the shaded region (shifted for clarity).}
     \label{fig:spectra_H13CN}
\end{figure}

\begin{figure}[ht]
\resizebox{\hsize}{!}
        {\includegraphics{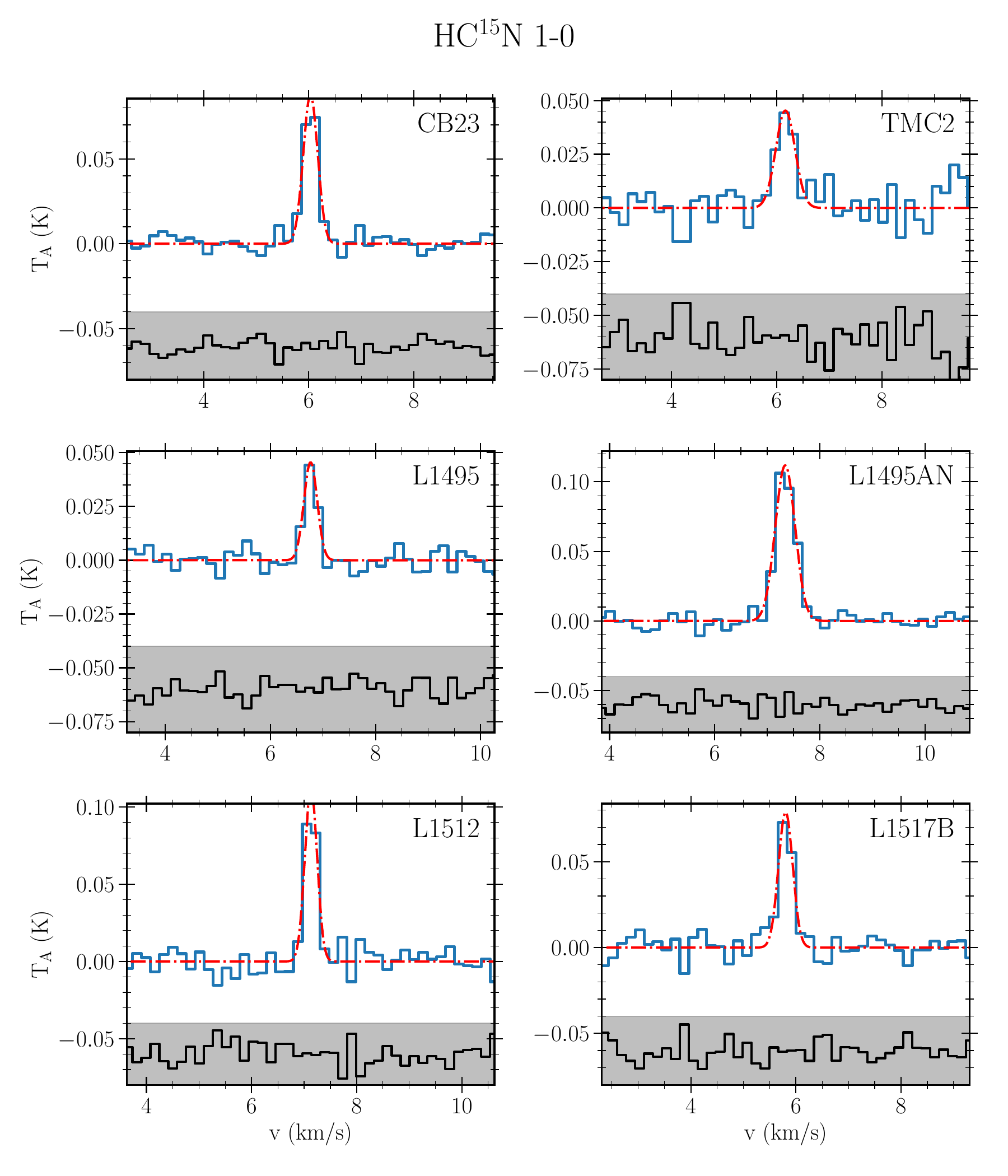}}
  \caption{Observed spectra for the 1--0 transition of HC${}^{15}$N for the six cores. The best Gaussian fit to the emission line is shown in red and residuals between the fit and the observed spectra are shown in the shaded region (shifted for clarity).}
     \label{fig:spectra_HC15N}
\end{figure}

\begin{figure}[ht]
\resizebox{\hsize}{!}
        {\includegraphics{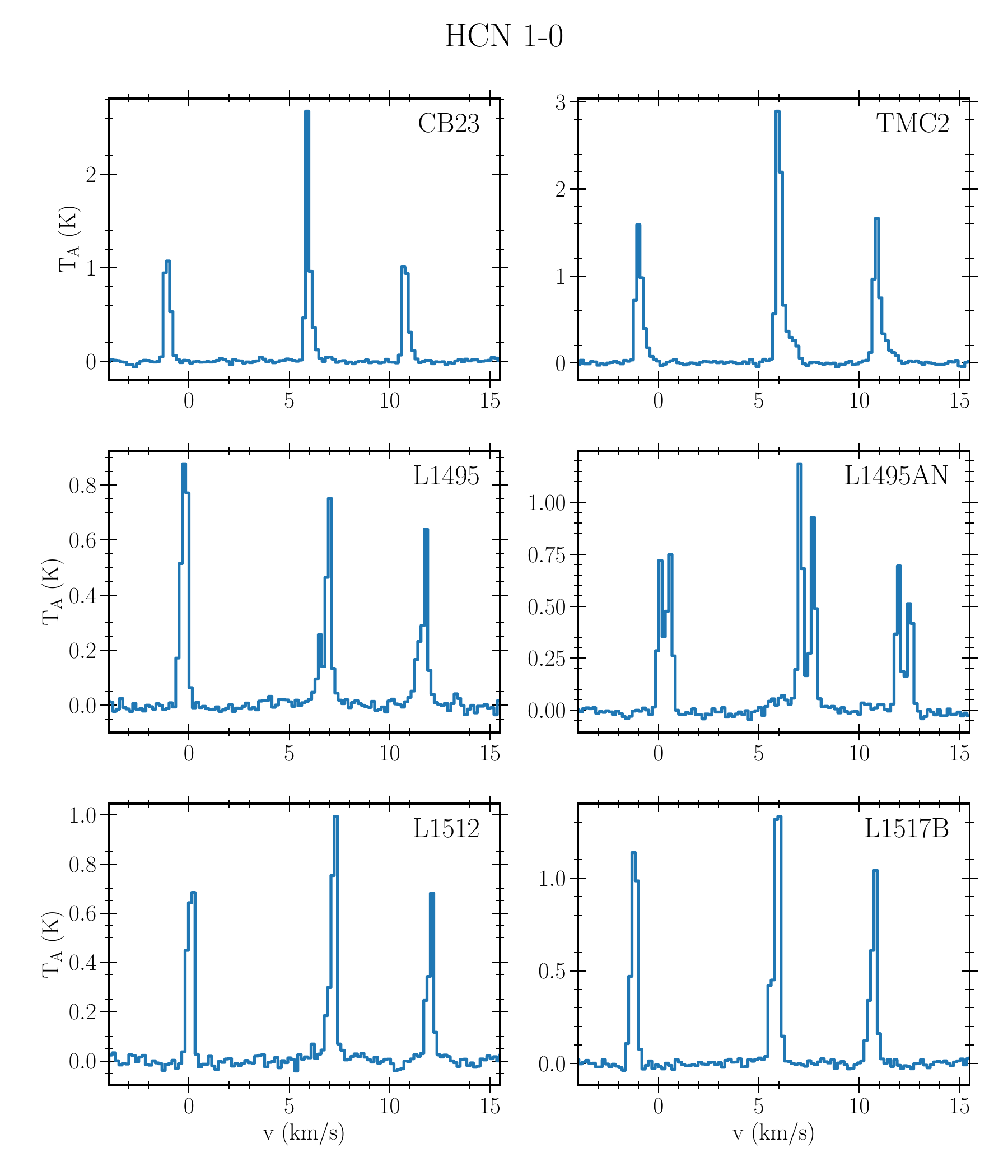}}
  \caption{Observed spectra for the 1--0 transition of HCN for the six cores.}
     \label{fig:spectra_HCN10}
\end{figure}

\begin{figure}[ht]
\resizebox{\hsize}{!}
        {\includegraphics{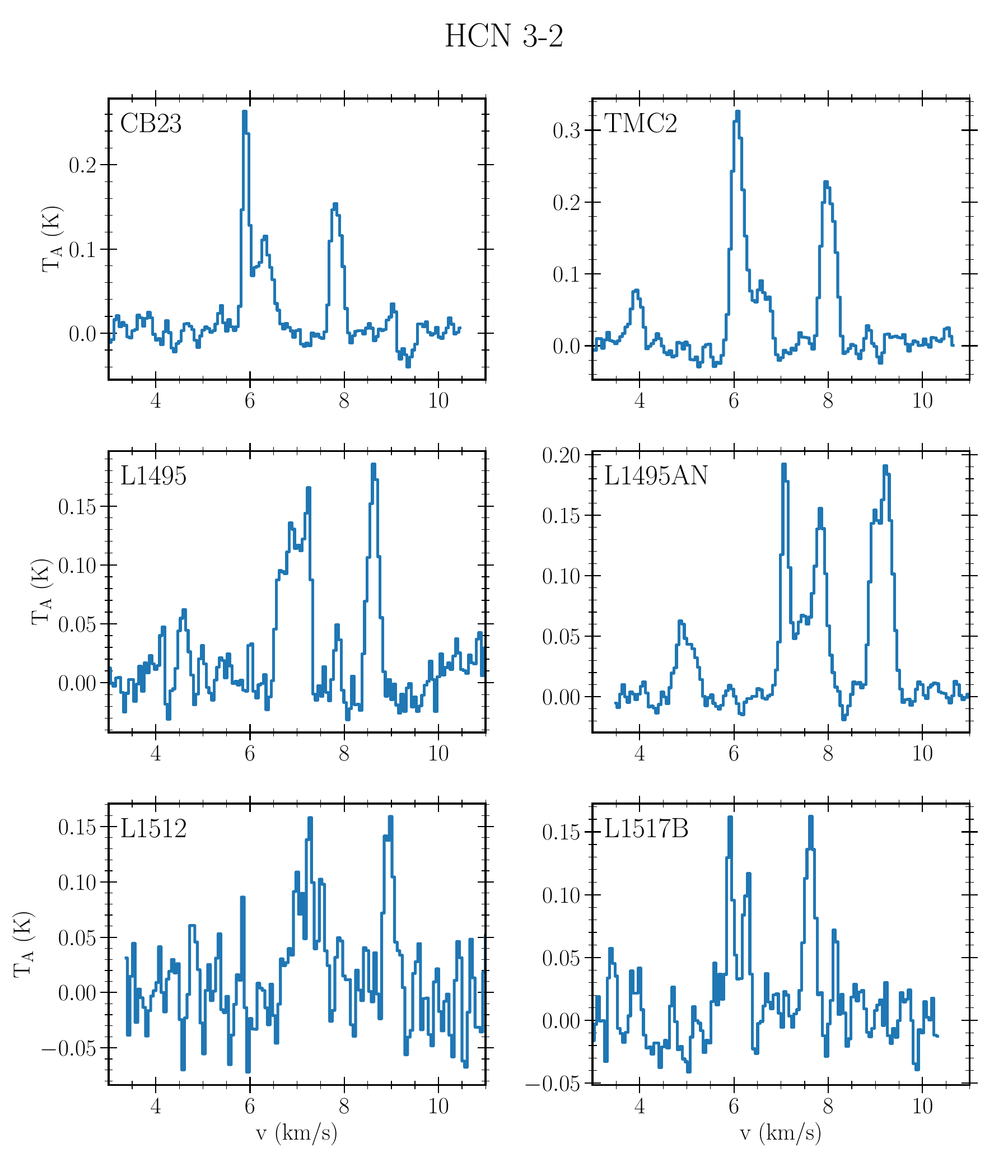}}
  \caption{Observed spectra for the 3--2 transition of HCN for the six cores.}
     \label{fig:spectra_HCN32}
\end{figure}

The observed spectra are shown in Figs. \ref{fig:spectra_H13CN}, \ref{fig:spectra_HC15N}, \ref{fig:spectra_HCN10}, and \ref{fig:spectra_HCN32}. For the 1--0 transitions of HCN, H$^{13}$CN, and HC$^{15}$N, a signal-to-noise ratio S/N > 10 was achieved toward all sources. For the 3--2 transition of HCN, the S/N varied significantly from source to source. This is due to the higher atmospheric absorption $\tau$ at 1~mm which makes observations of the 3--2 line more sensitive to local weather variations at Pico Veleta. 


\subsection{Spectral line fitting and the indirect method}
The nitrogen fractionation in HCN was first determined using the indirect method assuming a fixed $^{12}$C/$^{13}$C ratio of 68.
The approach involves fitting the hyperfine components of the 1-0 transition of H$^{13}$CN using the HFS fitting function in {\sc gildas} {\sc class} software.
The optical depth, column density, and excitation temperature of H$^{13}$CN are determined from the HFS fit. The best fit for each core is shown in Fig. \ref{fig:spectra_H13CN}, along with the residuals. For H${}^{13}$CN, the 1-0 transition is optically thin ($\tau < 0.4$ for the weakest HFS component) and the model fits the observed lines well as shown by the uniform residuals. The 1-0 transition of HC$^{15}$N is fitted using a single Gaussian line profile and the column density of HC$^{15}$N is determined by assuming the same excitation temperature as H${}^{13}$CN. Then, the HCN column density is estimated by multiplying the H$^{13}$CN column density by a factor of 68 from the canonical $^{12}$C/$^{13}$C ratio.
The results using this method are listed in Table \ref{table:indirect}.

\begin{table}[]
\centering\caption{Parameters for the HFS fit for H$^{13}$CN. $\tau_\mathrm{center}$ refers to the central HFS component at 86.3402~GHz, while $\tau_\mathrm{left}$ refers to the weakest component at 86.3423~GHz.}     
\label{table:hfs_fit}
\centering          
\begin{tabular}{c c c c}
\hline      
\hline      
Source & T$_\mathrm{ex}$ (K) & $\tau_\mathrm{center}$ & $\tau_\mathrm{left}$  \\
\hline      \noalign{\smallskip}
CB23      &    3.1$\pm$0.1       &     1.2$\pm$0.2 &     0.2$\pm$0.1    \\
TMC2      &     3.1$\pm$0.3    &     0.7$\pm$0.3    &     0.1$\pm$0.1  \\
L1495      &   3.3$\pm$0.3       &      0.4$\pm$0.2 &     0.1$\pm$0.1   \\
L1495AN      &    3.3$\pm$0.3      &     0.9$\pm$0.2 &     0.2$\pm$0.1   \\
L1512      &     3.3$\pm$0.1    &     1.1$\pm$0.2 &     0.2$\pm$0.1  \\
L1517B      &     3.1$\pm$0.1    &     1.6$\pm$0.3  &     0.3$\pm$0.1\\
\hline      \noalign{\smallskip}
\end{tabular}
\end{table}

\begin{table*}[]
\centering\caption{Estimated column densities derived using the indirect (double isotope) method assuming a fixed $^{12}$C/$^{13}$C ratio of 68. Uncertainties are propagated from the uncertainties on the fit and assume a fixed 10\% uncertainty on the flux.}     
\label{table:indirect}
\centering          
\begin{tabular}{c c c c c}
\hline      
\hline      
Source & $N$(H$^{13}$CN) ($10^{12}$~cm$^{-2}$) & $N$(HC$^{15}$N) ($10^{11}$~cm$^{-2}$) & $N$(HCN) ( $10^{13}$~cm$^{-2}$) & ${}^{14}$N/${}^{15}$N \\
\hline      \noalign{\smallskip}
CB23      &    1.4$\pm$0.2       &     2.4$\pm$0.6     &   9$\pm$2     & 390$\pm$59  \\
TMC2      &     1.1$\pm$0.4    &     1.6$\pm$0.7     &      7$\pm$3  &  468$\pm$123  \\
L1495      &   6$\pm$2       &      0.8$\pm$0.3    &    4$\pm$1    &  488$\pm$82  \\
L1495AN      &    1.6$\pm$0.3      &     3.3$\pm$0.8     &   11$\pm$2     &  340$\pm$53 \\
L1512      &     1.4$\pm$0.2    &     2.1$\pm$0.4     &     10$\pm$1   &  455$\pm$71  \\
L1517B      &     1.9$\pm$0.3    &     2.4$\pm$0.5     &    13$\pm$2    &  540$\pm$92 \\
\hline      \noalign{\smallskip}
\end{tabular}
\end{table*}

\subsection{Spectral modeling with {\sc LOC} and the direct method}
To model the observed spectra, we used the NLTE line radiative transfer code {\sc loc} \citep{2020A&A...644A.151J}.
Collisional rates for HCN were sourced from the LAMDA database \citep{2005A&A...432..369S}, relying on data from the CDMS database \citep{2001A&A...370L..49M, 2016JMoSp.327...95E} compiled from the studies of \citet{2002ZNatA..57..669A, 2004ZNatA..59..861F, 2005JMoSp.233..280C, 2005ApJS..159..181C, 2010MNRAS.406.2488D, 2013JChPh.139v4301D, 2017ApJ...836...30G}.

\begin{figure*}[ht]
\resizebox{\hsize}{!}
        {\includegraphics{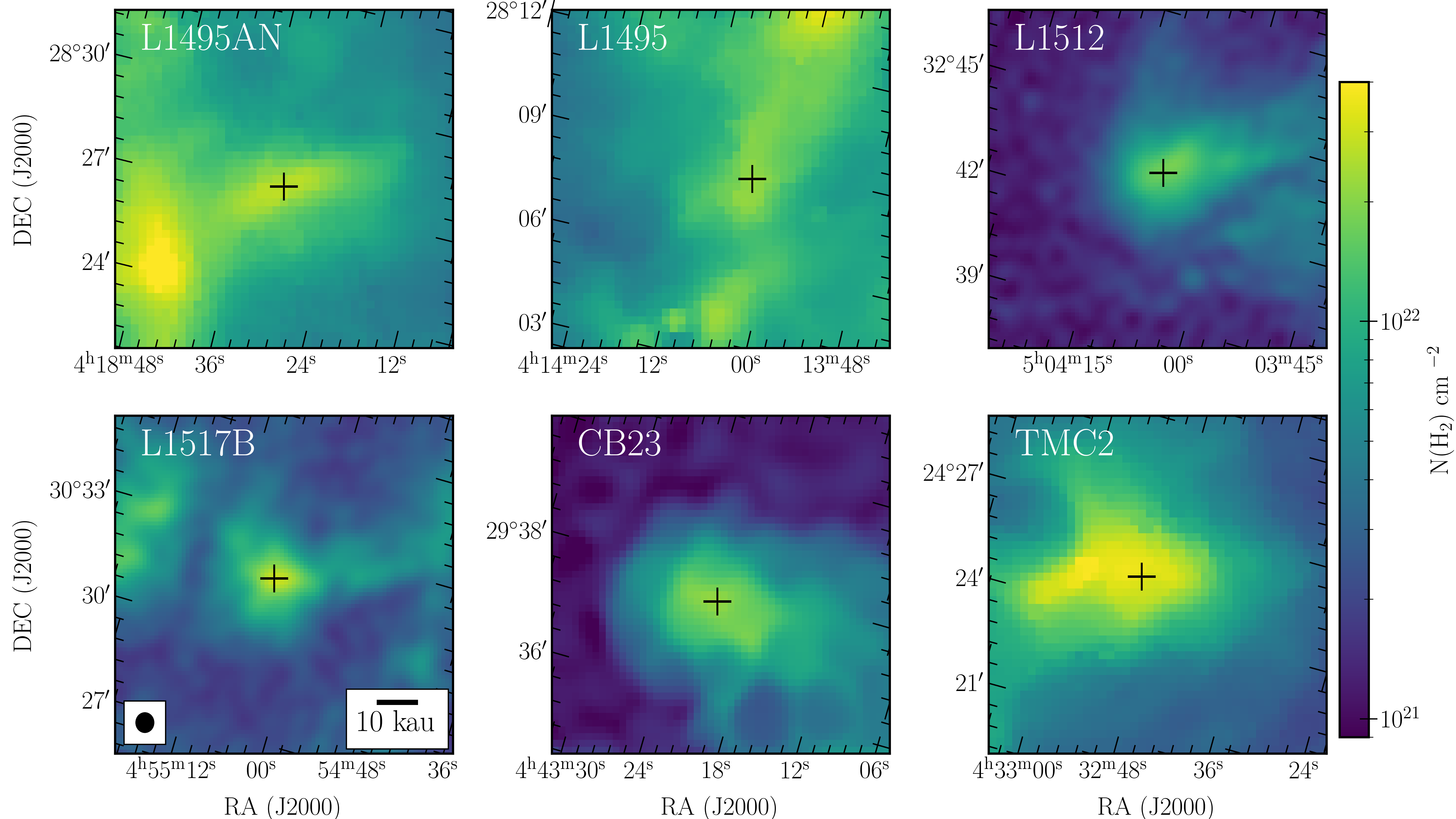}}
  \caption{Column density maps of H$_2$ derived from Herschel/SPIRE maps. The black cross marks the center of the core. The beam size is shown for the largest beam at $500~\mu$m and identical for all the maps. A distance of 140~pc is assumed for all sources.}
     \label{fig:cores}
\end{figure*}

For a direct determination of the HCN, H$^{13}$CN, and HC$^{15}$N column densities, we performed 1D radiative transfer modeling of each isotopolog.
For the radiative transfer modeling of the emission lines a source model is needed. To derive this model, we used \emph{Herschel}/SPIRE continuum maps of the cores at 250~$\mu$m, 350~$\mu$m, and 500~$\mu$m \citep{2010A&A...518L...3G}. The maps were obtained from the pipeline images available in the Herschel Science Archive \footnote{\url{http://archives.esac.esa.int/hsa/whsa/}}. From the three maps, the column density of molecular hydrogen $N$(H$_2$) was derived following the method of \citet{2016A&A...592L..11S}. Briefly, each map was smoothed to the resolution of the 500~$\mu$m image ($\sim$40$^{''}$) and resampled to the same grid. Next, a modified blackbody function with the dust emissivity spectral index $\beta = 2$ was fitted to each pixel \citep{2010ApJ...708..127S}. For the dust emissivity, we adopted the value $\kappa_{250\mu\mathrm{m}} = 0.1~\mathrm{cm}^{2} \mathrm{g}^{-1}$ from \citet{1983QJRAS..24..267H}. The column density maps derived for the six cores are shown in Fig. \ref{fig:cores}.

The density structure is fitted using a Plummer profile following \citet{2011A&A...529L...6A}: 
\begin{equation}
    \rho(r) = \frac{\rho_c}{[1 + (r/R_\mathrm{flat})^2]^{p/2}},
\end{equation}
where $\rho_c$ is the central density, $R_\mathrm{flat}$ is the characteristic radius of the inner flat region, and $p$ characterizes the slope of the density profile. The derived central densities $n(\mathrm{H}_2)$ lie in the range $(5-20)\times10^{4}$~cm$^{-3}$ and are comparable to the values presented in \citet{2005ApJ...619..379C} for a subset of the cores.

\begin{figure}[ht]
\resizebox{\hsize}{!}
        {\includegraphics{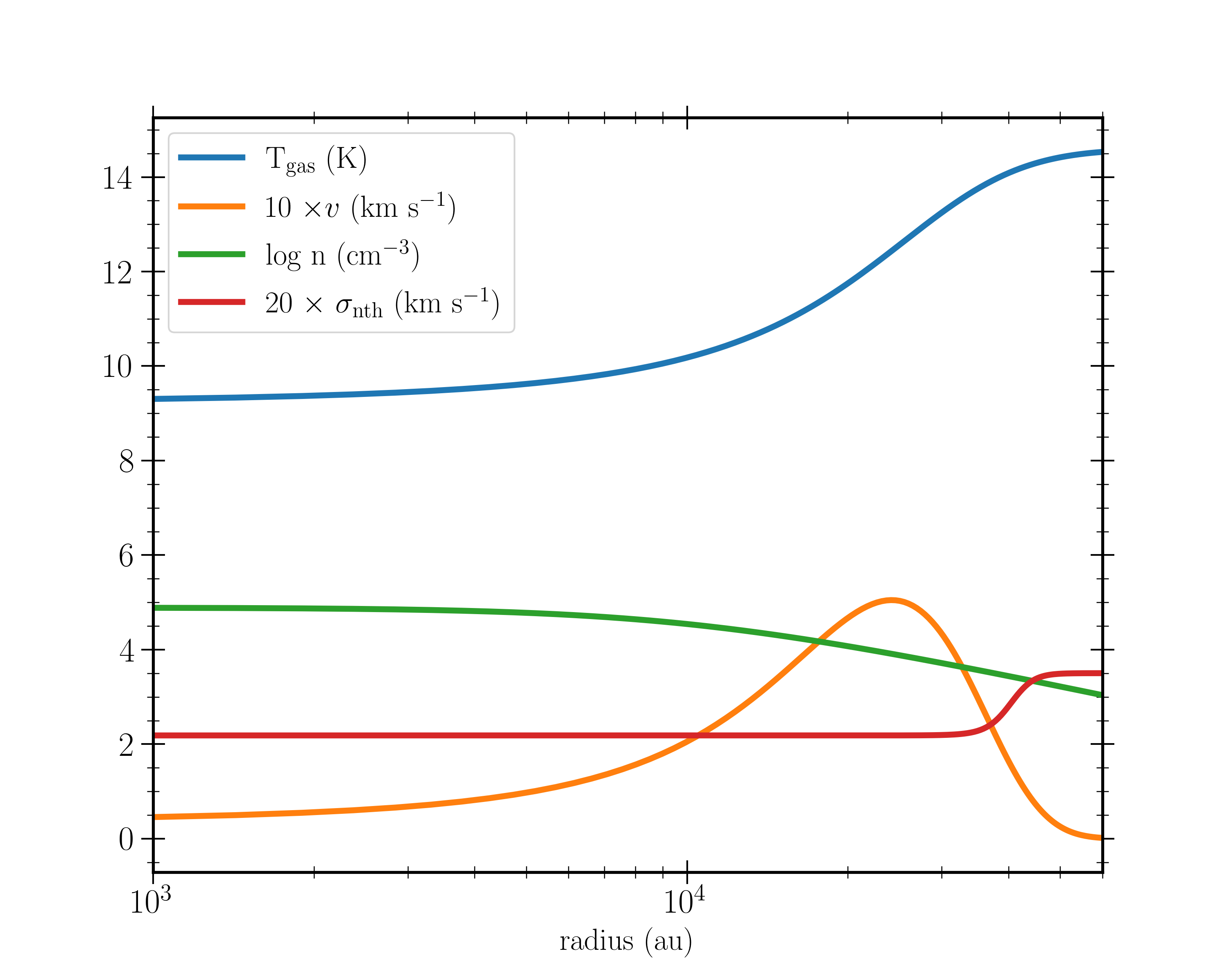}}
  \caption{Physical structure of TMC2 according to the best-fit model. The velocities are scaled for clarity.}
     \label{fig:TMC2_core}
\end{figure}

For the radial temperature profile, the non-thermal line broadening, and the (infall) velocity, we adopted the parametric functions presented in \citet{2018A&A...615A..52M}. 
The non-thermal velocity dispersion $\sigma$ is defined as:
\begin{equation}
    \sigma_\mathrm{nth} = \sigma_\mathrm{c} + \frac{\sigma_\mathrm{ext} - \sigma_\mathrm{c}}{\pi} \Big[\frac{\pi}{2} + \tanh (\frac{r-r_\mathrm{j}}{\Delta r_\mathrm{j}})\Big] \quad,
\end{equation}
where $\sigma_\mathrm{c}$ and $\sigma_\mathrm{ext}$ are the non-thermal broadening at the core center and core edge, respectively, and $r_\mathrm{j}$ and $\Delta r_\mathrm{j}$ define the position and the width of the transition between the two boundary values.
The collapse velocity profile is presented by a Gaussian profile:
\begin{equation} \label{eq:vel}
    \mathrm{v} = \mathrm{v}_\mathrm{c} \exp[-(r-r_\mathrm{v})^{2}/2\Delta r_\mathrm{v}^{2}], \quad
\end{equation}
where $v_\mathrm{c}$ denotes the peak collapse velocity, $r_\mathrm{v}$ denotes the position of the peak collapse velocity, and $\Delta r_\mathrm{v}$ defines the width of the collapsing region.
The temperature structure is defined as:
\begin{equation}
    T = T_\mathrm{in} + \frac{T_\mathrm{out}-T_\mathrm{in}}{2} [1 + \tanh (\frac{r - r_\mathrm{T}}{\Delta r_\mathrm{T}})] \quad,
\end{equation}
where $T_\mathrm{in}$ and $T_\mathrm{out}$ denote the central and edge temperature, respectively, and $r_\mathrm{T}$ and $\Delta r_\mathrm{T}$ define the position and the width of the transition between the two boundary values. Figure \ref{fig:TMC2_core} shows an example of the physical core structure derived  for TMC2 based on these profiles. We note that the dust temperature can also be derived from the \emph{Herschel}/SPIRE fitting performed to determine the densities, displaying a good agreement with the gas temperatures derived in the outer part of the core. We chose not to use these for the temperature profiles when fitting because of the low spatial resolution, which does not resolve the lower temperatures in the center of the cores.

Because the cores have relatively low central densities ($\lesssim 10^{5}$ cm~$^{-3}$) and are not as evolved as, for instance, L1544, we also ran models with a constant velocity throughout the core. For L1495, this approach was found to give a better result than the collapsing velocity profile and is therefore used instead. In addition to the Gaussian velocity profile in Eq. \ref{eq:vel}, a number of different profiles (e.g., log-normal and Weibul) were also tested, however these did not increase the quality of the fit.

The abundance of HCN is assumed to follow a step-function with an inner and outer abundance of HCN and a variable step radius. Each parameter in the above parameterizations is free in the subsequent modeling of the HCN lines.
The spectral line modeling is optimized using the Markov chain Monte Carlo (MCMC) code {\sc emcee} \citep{2013PASP..125..306F}.
The complete process for the determination of the $^{14}$N/$^{15}$N ratio with the direct method is described below. 

First, the density structure of the core is derived from the \emph{Herschel}/SPIRE maps. Then the 1-0 and 3-2 transitions of HCN are fitted simultaneously using the MCMC module combined with {\sc loc}. The intensities of the modeled spectra are compared bin-by-bin to the observed spectra and the $\chi^2$ is minimized. A total of 300 walkers and 500 steps are used for the MCMC chain. The first 200 steps are discarded as burn-in. Following this, a visual inspection of the median parameters from the MCMC chain is done to determine whether a satisfactory solution was found. Once the HCN transitions have been fitted, the 1-0 transitions of H$^{13}$CN and HC$^{15}$N are fitted independently using the same physical source model derived from the HCN optimization. Hence, the only free parameters are the abundances of the $^{13}$C and $^{15}$N isotopologs. This approach assumes that the three isotopologs are co-spatial, which is a common assumption in fractionation studies, but it should be noted that some local effects may arise from isotope-selective photodissociation. However, from the single-pointing data presented here, toward the dust continuum peak of each source, we do not expect a substantial impact on the results in this work.

\begin{figure}[ht]
\resizebox{\hsize}{!}
        {\includegraphics{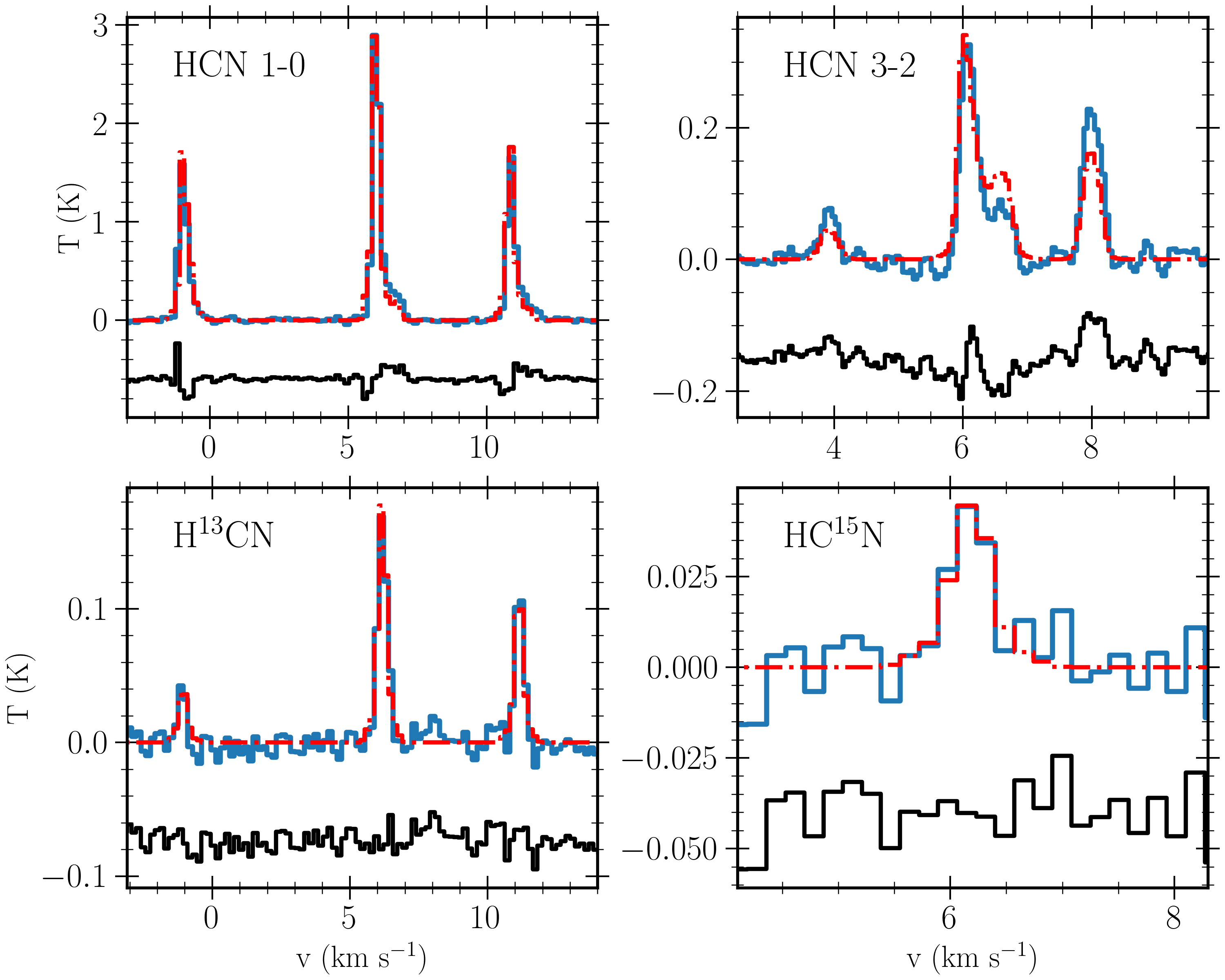}}
  \caption{Comparison between the modeled and observed spectra for the three isotopologs toward TMC2. Each panel shows the 1D spectral model in red and the observed spectrum in blue. Residuals between the model and the data are shown in black, offset for clarity.}
     \label{fig:model_TMC2}
\end{figure}

\begin{figure}[ht]
\resizebox{\hsize}{!}
        {\includegraphics{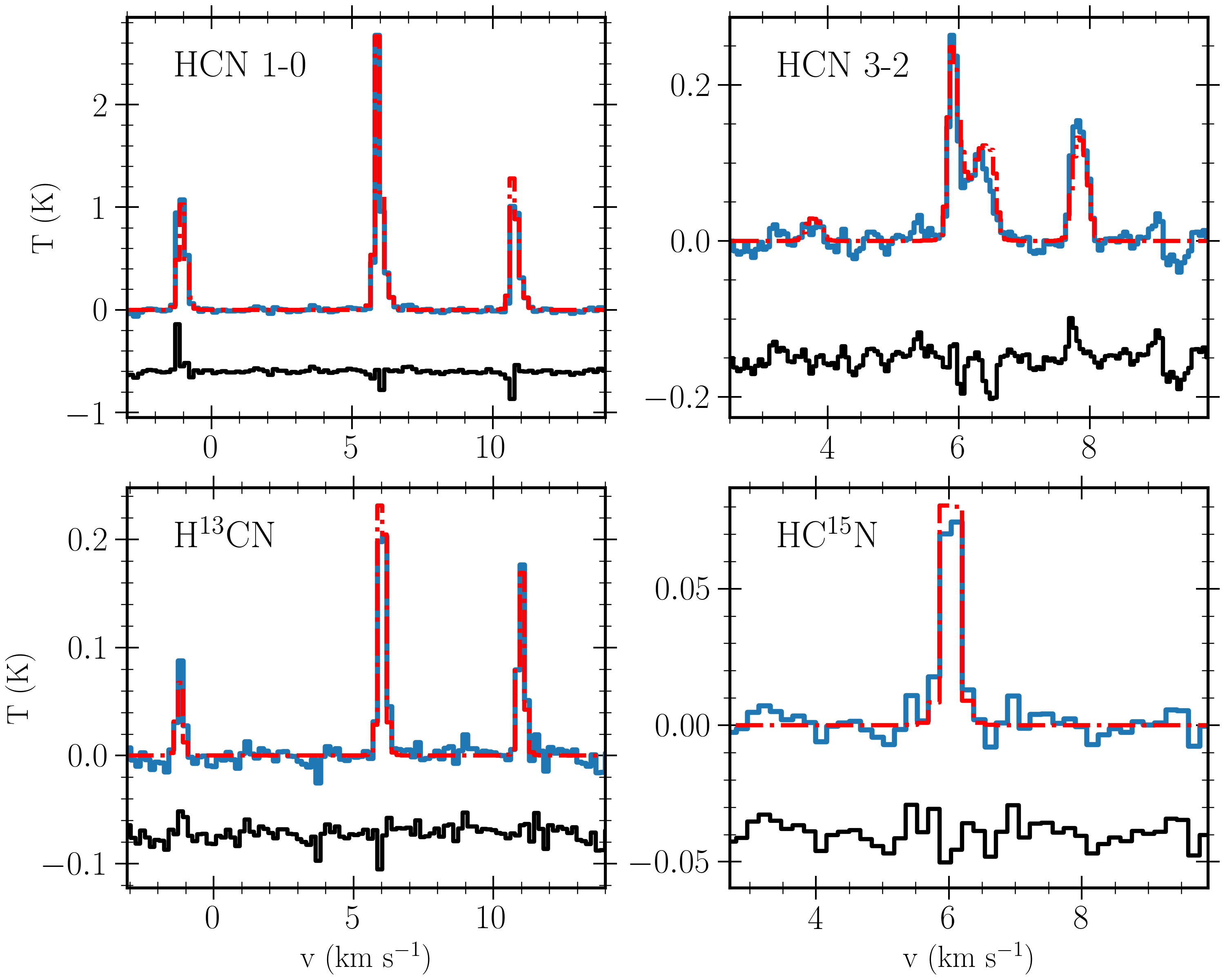}}
  \caption{Same as Fig. \ref{fig:model_TMC2}, for CB23.}
     \label{fig:model_CB23}
\end{figure}

\begin{figure}[ht]
\resizebox{\hsize}{!}
        {\includegraphics{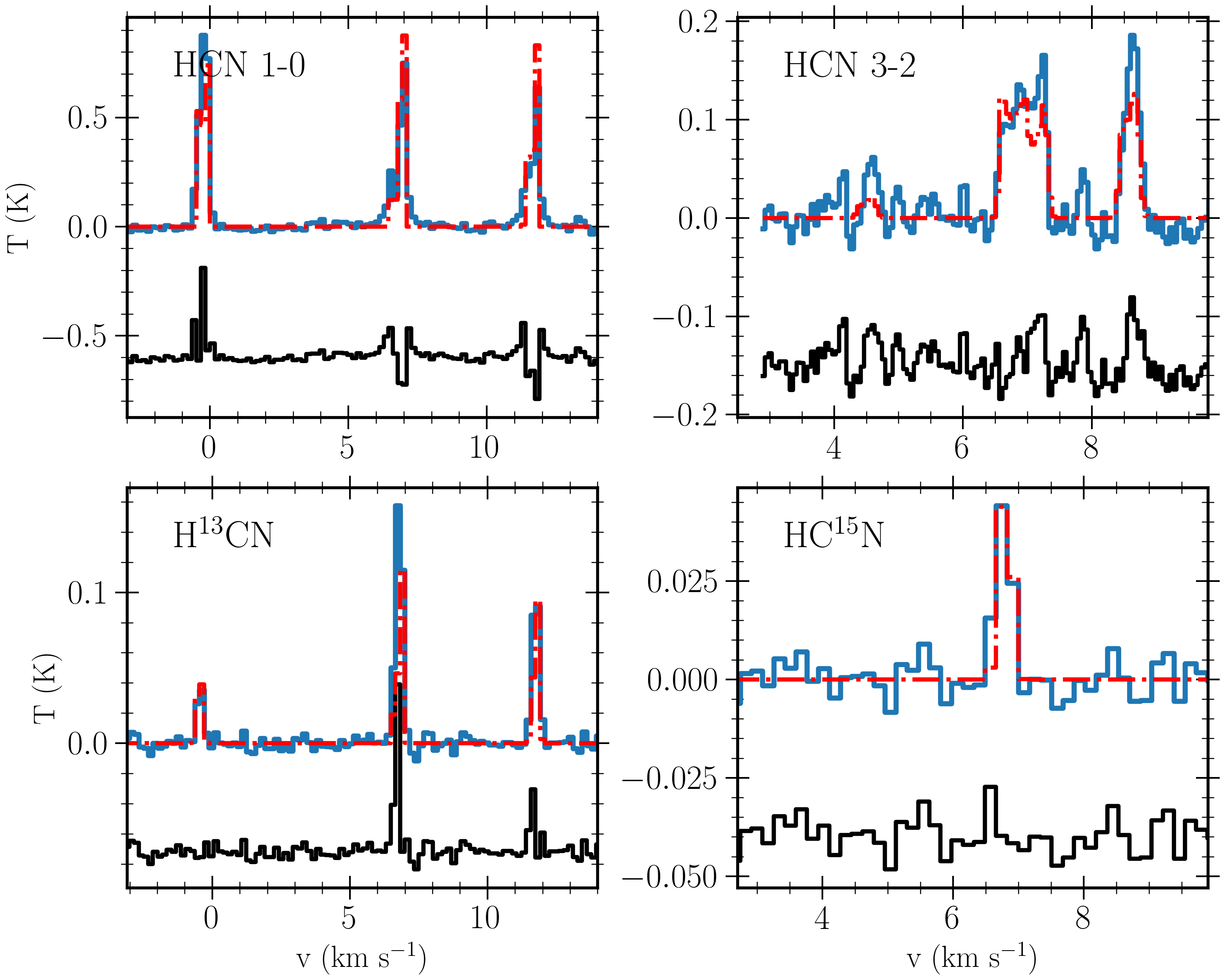}}
  \caption{Same as Fig. \ref{fig:model_TMC2}, for L1495.}
     \label{fig:model_L1495}
\end{figure}

\begin{figure}[ht]
\resizebox{\hsize}{!}
        {\includegraphics{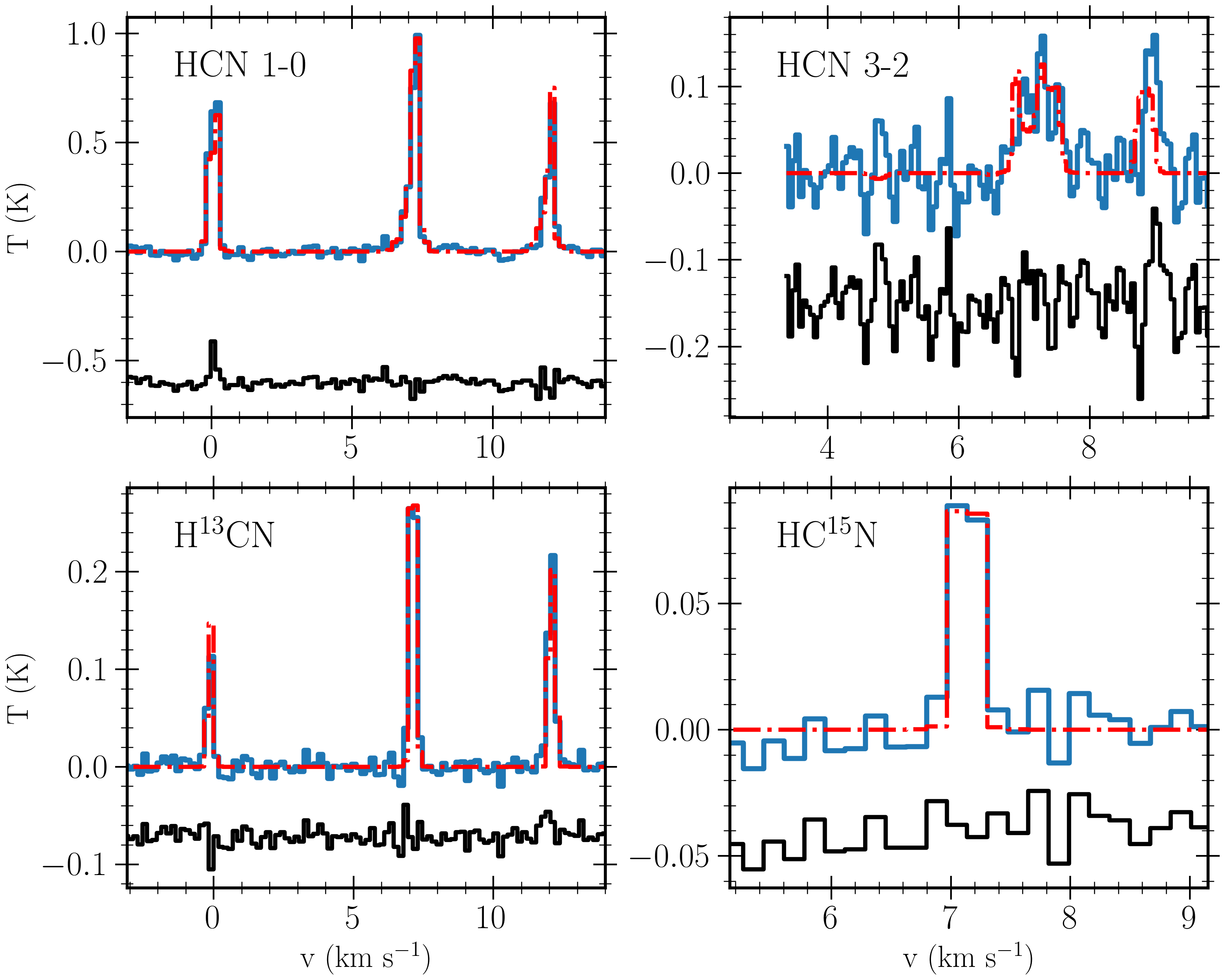}}
  \caption{Same as Fig. \ref{fig:model_TMC2}, for L1512.}
     \label{fig:model_L1512}
\end{figure}

\begin{figure}[ht]
\resizebox{\hsize}{!}
        {\includegraphics{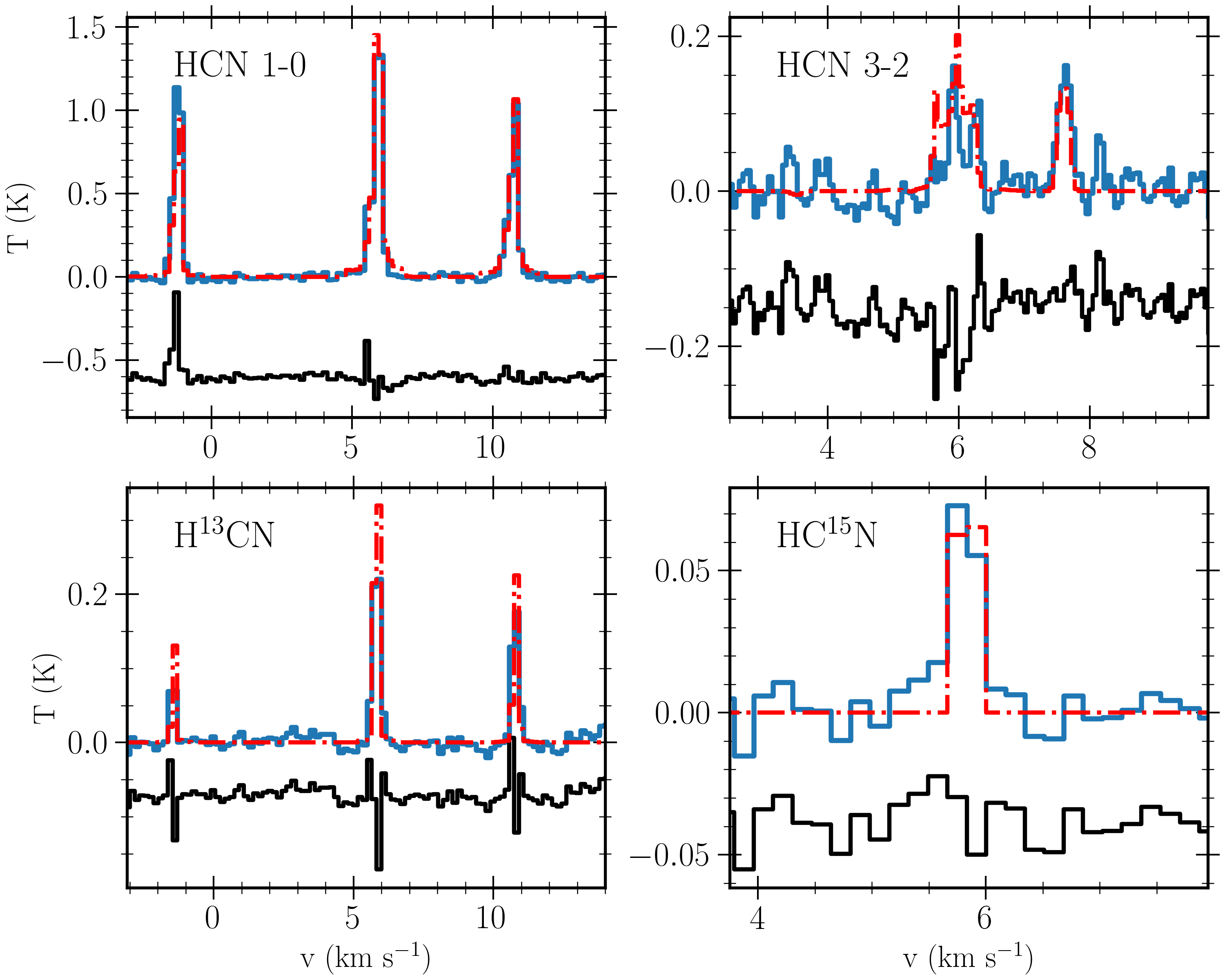}}
  \caption{Same as Fig. \ref{fig:model_TMC2}, for L1517B.}
     \label{fig:model_L1517B}
\end{figure}

Figures \ref{fig:model_TMC2}-\ref{fig:model_L1517B} show the median solution from the MCMC analysis for all sources except L1495AN, which was excluded as no satisfying solution was found. TMC2, L1512, and CB23 show the smallest residuals between the spectra and the spectral model across the four transitions, while the spectral models for L1495 and L1517B show larger residuals.


\begin{table*}[]
\centering\caption{Estimated column densities derived using the direct method. Uncertainties are propagated from the 25\% and 75\% percentiles of the MCMC chains.}     
\label{table:direct}
\centering          
\begin{tabular}{c c c c c c}
\hline    
\hline      
Source & $N$(H$^{13}$CN) ($10^{12}$~cm$^{-2}$) & $N$(HC$^{15}$N) ($10^{11}$~cm$^{-2}$) & $N$(HCN) ( $10^{13}$~cm$^{-2}$) & ${}^{14}$N/${}^{15}$N & ${}^{12}$C/${}^{13}$C \\
\hline      \noalign{\smallskip}
CB23      &    2.1$\pm$0.4       &     3.2$\pm$0.5     &   8$\pm$2     & 235$\pm$58 & 36$\pm10$  \\
TMC2      &     2.3$\pm$0.2    &     3.2$\pm$0.6     &      15$\pm$3  &  461$\pm$130 & 63$\pm15$ \\
L1512      &   7.9$\pm$0.9       &      6.5$\pm$0.8    &    16$\pm$4    &  251$\pm$76 & 20$\pm7$  \\
L1495      &     6.1$\pm$0.8    &     8$\pm$1     &     20$\pm$5   &  197$\pm$61 & 25$\pm$10  \\
L1517B      &     4.4$\pm$0.3    &     3.9$\pm$0.4     &    11$\pm$4    &  280$\pm$85 & 25$\pm$8  \\
\hline      \noalign{\smallskip}
\end{tabular}
\end{table*}

\begin{figure*}[ht]
\resizebox{\hsize}{!}
        {\includegraphics{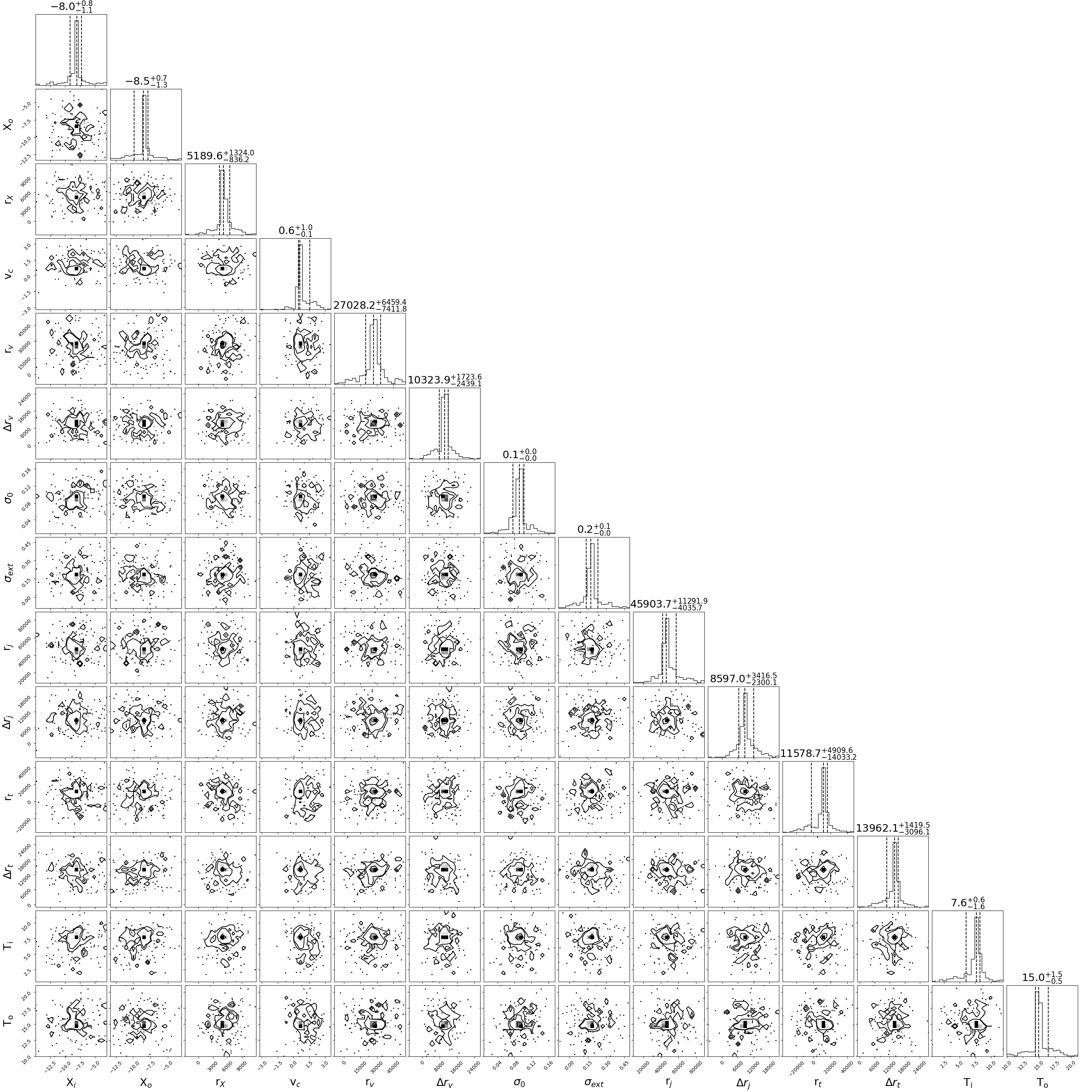}}
  \caption{Corner plot for the fitting of HCN and the physical structure of the core for TMC2.}
     \label{fig:corner}
\end{figure*}

\begin{figure}[ht]
\resizebox{\hsize}{!}
        {\includegraphics{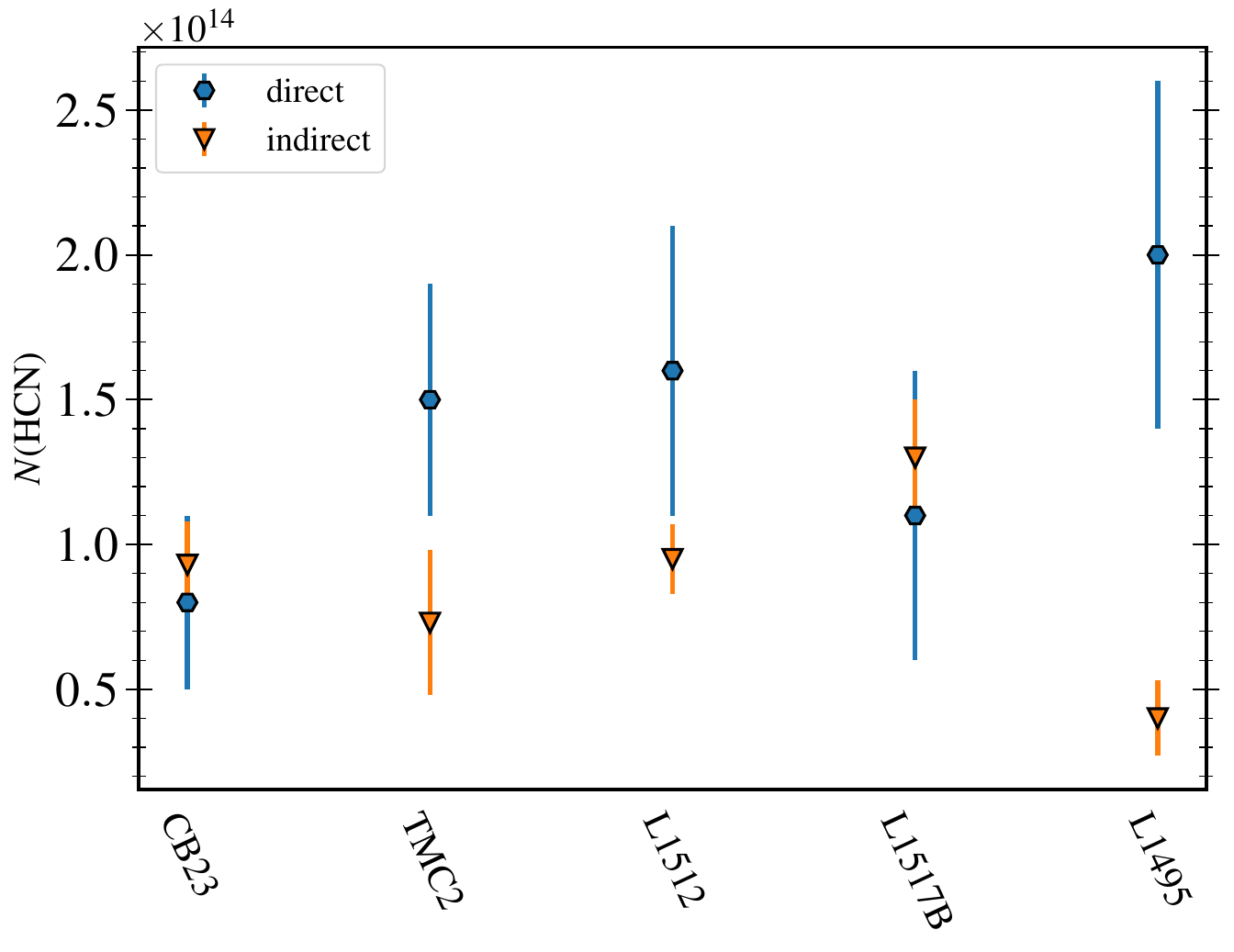}}
  \caption{Comparison between estimated HCN column densities using the direct and indirect methods. In general, the column densities are underestimated using the indirect method, except for CB23 and L1517B where the results of the two methods are within the uncertainties.}
     \label{fig:HCN_comparison}
\end{figure}

Corner plots showing the distribution of the parameter space explored by the workers are shown in Fig. \ref{fig:corner} for TMC2. Figures \ref{fig:corner_CB23}-\ref{fig:corner_L1517B} show the corner plots for the remaining sources. In the case of TMC2, the plot indicates that most parameters are well-constrained, with approximately Gaussian distributions around the median. The results of the direct method are listed in Table \ref{table:direct} and a comparison between the HCN column densities derived using the direct and indirect methods are presented in Fig. \ref{fig:HCN_comparison}. The direct method generally predicts higher HCN column densities than the indirect, corresponding to a lower $^{12}$C/$^{13}$C ratio.


 

\section{Discussion}  \label{sec:4}

\subsection{Comparison of the direct and indirect method}
Determining the abundance of HCN is often challenging due to its high optical thickness. This issue can be circumvented by using the indirect double isotope method, where a fixed $^{12}$C/$^{13}$C ratio is assumed. However, the extent to which the assumption of a fixed $^{12}$C/$^{13}$C ratio is reliable is unclear and may depend on both the type of source studied and the molecule in question.

In Fig. \ref{fig:Nfrac_NEW}, we present the $^{14}$N/$^{15}$N ratio of HCN for various evolutionary stages when using both the indirect double isotope method (blue) and the direct method (brown) for starless and pre-stellar cores in the literature as well as those presented in this work. A clear shift in the mean $^{14}$N/$^{15}$N ratio (dash-dotted lines) between the two methods is present. This shift is large enough to alter the interpretation of the current data, as discussed in the next section. The observed shift in the nitrogen fractionation is directly related to the $^{12}$C/$^{13}$C ratios in HCN, which are generally between 20-40, instead of the local ISM values of 68. Here, one exception is TMC2, which is consistent with the ISM value. The derived $^{12}$C/$^{13}$C ratios are generally consistent with those derived in other cold sources such as L1498 \citep[45$\pm$3,][]{2018A&A...615A..52M}, B1-b \citep[30$^{+7}_{-4}$,][]{2013A&A...560A...3D}, and the cold envelope of L483 \citep[$34\pm$10,][]{2019A&A...625A.147A}. 
If we assume a lower ratio of, for instance, $^{12}$C/$^{13}$C = 30 for all starless and pre-stellar cores, the shift between the two methods falls within the uncertainty of these measurements. This underlines the significance of the assumed $^{12}$C/$^{13}$C ratio and the weakness of the indirect method. 
With the data presented here and the results from other cold sources listed above, the assumption of a fixed $^{12}$C/$^{13}$C ratio in the starless and pre-stellar core stage appear not to be valid. This is also supported by the recent astrochemical modeling of the $^{12}$C/$^{13}$C ratio in HCN, which does not find that a fixed value of the $^{12}$C/$^{13}$C ratio is a good assumption \citep{2020A&A...640A..51C, 2020MNRAS.498.4663Ld}. These modeling efforts suggest that the ratio varies significantly over time and with physical conditions. \citet{2023A&A...678A.120S} modeled the fractionation ratios of HCN including both $^{13}$C and $^{15}$N in their network. By comparing the modeled abundances of HCN, HC$^{15}$N, and H$^{13}$CN, they demonstrated that the indirect method using $^{12}$C/$^{13}$C = 68 can lead to errors of an order of magnitude in the inferred $^{14}$N/$^{15}$N ratio. Their chemical model predicts $^{12}$C/$^{13}$C ratios below 50 in cold and dense environments ($T$=7~K, n$_\mathrm{H_2}$=10$^{6}$~cm$^{-3}$, $t$ > 10$^5$~yrs, private comm.), in agreement with the observational results for cold sources.

The indirect method is widely used at the protostellar and protoplanetary disk stages \citep[e.g.,][]{2014A&A...572A..24W, 2017ApJ...836...30G}, where the effects of the double isotope method is less studied. Currently, the $^{14}$N/$^{15}$N ratio of HCN have been estimated directly in one disk, TW Hydra \citep{2019A&A...632L..12H}, in which a disk-averaged $^{12}$C/$^{13}$C ratio of 84$\pm$4 and $^{14}$N/$^{15}$N ratio of 223$\pm$21 was found. Again, the carbon fractionation ratio differs from 68 and may induce systematic errors if the indirect double isotope method is used in protoplanetary disks. 
More direct determinations of the HCN abundance and the $^{12}$C/$^{13}$C and $^{14}$N/$^{15}$N ratios toward protostars and protoplanetary disks are needed to determine the robustness of the indirect method in these sources. 

\subsection{Inheritance or processing of HCN during star formation}
Nitrogen fractionation can serve as a powerful tracer of the chemical evolution during star- and planet formation. This requires robust determinations of the different isotopic abundances at different stages of the star and planet formation process to trace how the fractionation evolves with time. 
Due to the possible systematic errors when using the indirect method as highlighted in the previous section, care should be taken when comparing nitrogen fractionation ratios across evolutionary stages using different methods.

\begin{figure*}[ht]
\resizebox{\hsize}{!}
        {\includegraphics{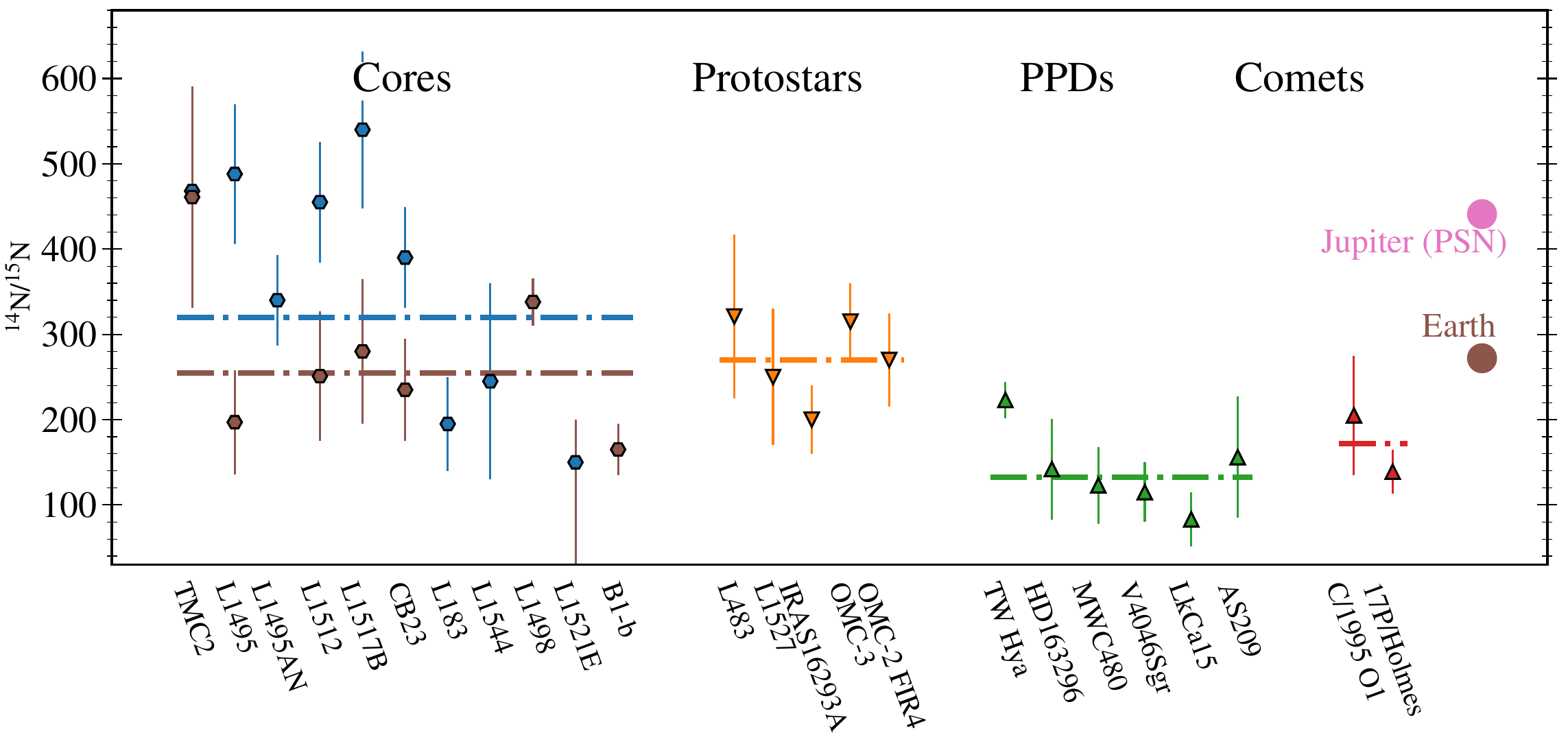}}
  \caption{Overview of reported $^{14}$N/$^{15}$N ratios in HCN at various evolutionary stages from cores to disks. For cores, the indirect measurements are indicated in blue and the direct in brown. The majority of the measurements for both protostars and protoplanetary disks (PPDs) are derived using the indirect double isotope method and hence assume a fixed $^{12}$C/$^{13}$C ratio of 68. The dashed-dotted lines show the weighted average of each population.}
     \label{fig:Nfrac_NEW}
\end{figure*}

The majority of the measurements presented in Fig. \ref{fig:Nfrac_NEW} are derived using the indirect method. If we only consider isotopic ratios derived using the direct method, then the nitrogen fractionation ratio in HCN remains almost constant within the reported errors. This is suggesting that no significant processing of HCN occurs between the different evolutionary stages, which influences the nitrogen fractionation. However, the picture is different if we include all measurements from indirect measurements. Here, a lower $^{14}$N/$^{15}$N ratio is seen at the protoplanetary disk stage compared to the starless and pre-stellar core stage.
Based on the current data, we cannot  determine whether HCN is processed during the collapse of the core to form the protostar and protoplanetary disk. 
\subsection{Challenges of the direct method}
The results presented here, along with the recent astrochemical modeling of carbon and nitrogen fractionation \citep[e.g.,][]{2023A&A...678A.120S} suggest that the indirect method is unreliable and should be used cautiously. However, using radiative transfer modeling to provide direct estimates of the HCN abundances is also challenging, as this work shows. The uncertainty on the source structure results in a large number of free parameters. The models presented here include up to 14 parameters when fitting the HCN 1-0 and 3-2 hyperfine lines. This number can be reduced if better constraints on the physical structure of the cores can be obtained, for example, through additional continuum observations with a broader wavelength coverage and higher spatial resolution. Furthermore, the assumption of a 1D core structure is also in itself a limitation, as shown by chemical segregation in L1544 \citep[e.g.,][]{2016A&A...592L..11S} and recent astrochemical modeling \citep[e.g.,][]{2023MNRAS.518.4839P, 2023A&A...675A..34J}. Another challenge when modeling HCN arises from the hyperfine anomalies reported for a wide number of different sources \citep{2016MNRAS.459.2882M}. The cause of these anomalies is not yet understood and they make the spectral fitting challenging.
Because of the challenges listed above, the approach of observing H$^{13}$C$^{15}$N, H$^{13}$CN, and HC$^{15}$N should also be considered in future works when sufficient S/N can be achieved, as this method does not rely on a well-constrained source structure. However, robust detections of the weaker H$^{13}$C$^{15}$N transition in the sources presented here would require decreasing the root mean square (RMS) noise level by a factor of $\sim$10, meaning a increased integration time by a factor of $\sim$100. Measurements of H$^{13}$C$^{15}$N are therefore likely to be reserved for sources with higher column densities or deep studies of individual sources.

\section{Conclusions}  \label{sec:5}

This paper presents IRAM 30m observations of HCN, H$^{13}$CN, and HC$^{15}$N toward six starless and pre-stellar cores. The nitrogen fractionation ratio $^{14}$N/$^{15}$N is determined using first the indirect double isotope method and later directly by fitting the spectra with NLTE radiative transfer models of the cores to determine the molecular abundances. The latter method is challenging and only successful for five of the six cores.

With the results of the indirect method, an apparent trend in the $^{14}$N/$^{15}$N ratio of HCN from starless cores to protostars and protoplantery disks is present, with higher values around $^{14}$N/$^{15}$N $\sim$ 300-550 in the earlier core phases and $^{14}$N/$^{15}$N $<$ 250 in the later protoplanetary disk phase. However, this trend is not present when we instead use the direct method to determine both $^{14}$N/$^{15}$N and $^{12}$C/$^{13}$C ratios directly. The difference between the two methods can be attributed to a lower $^{12}$C/$^{13}$C ratio in HCN within the range 20-40 for most cores instead of the ISM values of 68. When the results of the direct method are used, the apparent evolutionary trend in the $^{14}$N/$^{15}$N ratio is eliminated, which could suggest that HCN is inherited from the earliest stages of star formation and does not experience significant processing. However, as many of the previous measurements are done using the indirect method, more direct measurements needs to be done in the protostellar and protoplanetary disk phases to test this statement. Since this work shows a substantial difference in the interpretation of the data when the direct and indirect methods are used, caution should be exercised while comparing direct and indirect data points in the search for evolutionary trends. 
Ultimately, the indirect method appears unreliable and future work should aim to obtain direct estimates of the $^{14}$N/$^{15}$N ratio at all evolutionary stages to determine whether HCN is inherited or processed during the formation of stars and planets. This can be achieved either through radiative transfer modeling (as in this work) or by observing the double-substituted H$^{13}$C$^{15}$N (when feasible). While these direct methods are more challenging, they provide additional data on the carbon fractionation during star and planet formation, further improving our understanding of chemical evolution in the interstellar medium.

\begin{acknowledgements}
The authors wish to thank the referee for useful comments which helped improve the manuscript.
S.S.J., S.S., and K. G. wish to thank the Max Planck Society for the Max Planck Research Group funding. All other authors
affiliated to the MPE wish to thank the Max Planck Society for financial support.
The Tycho supercomputer hosted at the SCIENCE HPC center at the University of Copenhagen was used for supporting this work. We acknowledge project support by the Max Planck Computing and Data Facility.
 This paper makes use of {\sc matplotlib} \citep{hunter2007}, {\sc emcee} \citep{2013PASP..125..306F}, and {\sc scipy} \citep{2020SciPy-NMeth}.
\end{acknowledgements}

%
%

\bibliographystyle{aa}
\bibliography{n.bib}

\begin{thebibliography}{56}
\expandafter\ifx\csname natexlab\endcsname\relax\def\natexlab#1{#1}\fi

\bibitem[{{Ag{\'u}ndez} {et~al.}(2019){Ag{\'u}ndez}, {Marcelino}, {Cernicharo},
  {Roueff}, \& {Tafalla}}]{2019A&A...625A.147A}
{Ag{\'u}ndez}, M., {Marcelino}, N., {Cernicharo}, J., {Roueff}, E., \&
  {Tafalla}, M. 2019, \aap, 625, A147

\bibitem[{{Ahrens} {et~al.}(2002){Ahrens}, {Lewen}, {Takano}, {Winnewisser},
  {Urban}, {Negirev}, \& {Koroliev}}]{2002ZNatA..57..669A}
{Ahrens}, V., {Lewen}, F., {Takano}, S., {et~al.} 2002, Zeitschrift
  Naturforschung Teil A, 57, 669

\bibitem[{{Arzoumanian} {et~al.}(2011){Arzoumanian}, {Andr{\'e}}, {Didelon},
  {K{\"o}nyves}, {Schneider}, {Men'shchikov}, {Sousbie}, {Zavagno}, {Bontemps},
  {di Francesco}, {Griffin}, {Hennemann}, {Hill}, {Kirk}, {Martin}, {Minier},
  {Molinari}, {Motte}, {Peretto}, {Pezzuto}, {Spinoglio}, {Ward-Thompson},
  {White}, \& {Wilson}}]{2011A&A...529L...6A}
{Arzoumanian}, D., {Andr{\'e}}, P., {Didelon}, P., {et~al.} 2011, \aap, 529, L6

\bibitem[{{Bizzocchi} {et~al.}(2013){Bizzocchi}, {Caselli}, {Leonardo}, \&
  {Dore}}]{2013A&A...555A.109B}
{Bizzocchi}, L., {Caselli}, P., {Leonardo}, E., \& {Dore}, L. 2013, \aap, 555,
  A109

\bibitem[{{Bockel{\'e}e-Morvan} {et~al.}(2008){Bockel{\'e}e-Morvan}, {Biver},
  {Jehin}, {Cochran}, {Wiesemeyer}, {Manfroid}, {Hutsem{\'e}kers}, {Arpigny},
  {Boissier}, {Cochran}, {Colom}, {Crovisier}, {Milutinovic}, {Moreno},
  {Prochaska}, {Ramirez}, {Schulz}, \& {Zucconi}}]{2008ApJ...679L..49B}
{Bockel{\'e}e-Morvan}, D., {Biver}, N., {Jehin}, E., {et~al.} 2008, \apjl, 679,
  L49

\bibitem[{{Caselli} \& {Ceccarelli}(2012)}]{2012A&ARv..20...56C}
{Caselli}, P. \& {Ceccarelli}, C. 2012, \aapr, 20, 56

\bibitem[{{Cazzoli} \& {Puzzarini}(2005)}]{2005JMoSp.233..280C}
{Cazzoli}, G. \& {Puzzarini}, C. 2005, Journal of Molecular Spectroscopy, 233,
  280

\bibitem[{{Cazzoli} {et~al.}(2005){Cazzoli}, {Puzzarini}, \&
  {Gauss}}]{2005ApJS..159..181C}
{Cazzoli}, G., {Puzzarini}, C., \& {Gauss}, J. 2005, \apjs, 159, 181

\bibitem[{{Cleeves} {et~al.}(2014){Cleeves}, {Bergin}, {Alexander}, {Du},
  {Graninger}, {{\"O}berg}, \& {Harries}}]{2014Sci...345.1590C}
{Cleeves}, L.~I., {Bergin}, E.~A., {Alexander}, C. M.~O.~D., {et~al.} 2014,
  Science, 345, 1590

\bibitem[{{Colzi} {et~al.}(2018{\natexlab{a}}){Colzi}, {Fontani}, {Caselli},
  {Ceccarelli}, {Hily-Blant}, \& {Bizzocchi}}]{2018A&A...609A.129C}
{Colzi}, L., {Fontani}, F., {Caselli}, P., {et~al.} 2018{\natexlab{a}}, \aap,
  609, A129

\bibitem[{{Colzi} {et~al.}(2018{\natexlab{b}}){Colzi}, {Fontani}, {Rivilla},
  {S{\'a}nchez-Monge}, {Testi}, {Beltr{\'a}n}, \&
  {Caselli}}]{2018MNRAS.478.3693C}
{Colzi}, L., {Fontani}, F., {Rivilla}, V.~M., {et~al.} 2018{\natexlab{b}},
  \mnras, 478, 3693

\bibitem[{{Colzi} {et~al.}(2020){Colzi}, {Sipil{\"a}}, {Roueff}, {Caselli}, \&
  {Fontani}}]{2020A&A...640A..51C}
{Colzi}, L., {Sipil{\"a}}, O., {Roueff}, E., {Caselli}, P., \& {Fontani}, F.
  2020, \aap, 640, A51

\bibitem[{{Crapsi} {et~al.}(2005){Crapsi}, {Caselli}, {Walmsley}, {Myers},
  {Tafalla}, {Lee}, \& {Bourke}}]{2005ApJ...619..379C}
{Crapsi}, A., {Caselli}, P., {Walmsley}, C.~M., {et~al.} 2005, \apj, 619, 379

\bibitem[{{Daniel} {et~al.}(2013){Daniel}, {G{\'e}rin}, {Roueff}, {Cernicharo},
  {Marcelino}, {Lique}, {Lis}, {Teyssier}, {Biver}, \&
  {Bockel{\'e}e-Morvan}}]{2013A&A...560A...3D}
{Daniel}, F., {G{\'e}rin}, M., {Roueff}, E., {et~al.} 2013, \aap, 560, A3

\bibitem[{{Denis-Alpizar} {et~al.}(2013){Denis-Alpizar}, {Kalugina},
  {Stoecklin}, {Vera}, \& {Lique}}]{2013JChPh.139v4301D}
{Denis-Alpizar}, O., {Kalugina}, Y., {Stoecklin}, T., {Vera}, M.~H., \&
  {Lique}, F. 2013, \jcp, 139, 224301

\bibitem[{{Dumouchel} {et~al.}(2010){Dumouchel}, {Faure}, \&
  {Lique}}]{2010MNRAS.406.2488D}
{Dumouchel}, F., {Faure}, A., \& {Lique}, F. 2010, \mnras, 406, 2488

\bibitem[{{Endres} {et~al.}(2016){Endres}, {Schlemmer}, {Schilke}, {Stutzki},
  \& {M{\"u}ller}}]{2016JMoSp.327...95E}
{Endres}, C.~P., {Schlemmer}, S., {Schilke}, P., {Stutzki}, J., \&
  {M{\"u}ller}, H. S.~P. 2016, Journal of Molecular Spectroscopy, 327, 95

\bibitem[{{Foreman-Mackey} {et~al.}(2013){Foreman-Mackey}, {Hogg}, {Lang}, \&
  {Goodman}}]{2013PASP..125..306F}
{Foreman-Mackey}, D., {Hogg}, D.~W., {Lang}, D., \& {Goodman}, J. 2013, \pasp,
  125, 306

\bibitem[{{Fouchet} {et~al.}(2004){Fouchet}, {Irwin}, {Parrish}, {Calcutt},
  {Taylor}, {Nixon}, \& {Owen}}]{2004Icar..172...50F}
{Fouchet}, T., {Irwin}, P. G.~J., {Parrish}, P., {et~al.} 2004, \icarus, 172,
  50

\bibitem[{{Fuchs} {et~al.}(2004){Fuchs}, {Bruenken}, {Fuchs}, {Thorwirth},
  {Ahrens}, {Lewen}, {Urban}, {Giesen}, \& {Winnewisser}}]{2004ZNatA..59..861F}
{Fuchs}, U., {Bruenken}, S., {Fuchs}, G.~W., {et~al.} 2004, Zeitschrift
  Naturforschung Teil A, 59, 861

\bibitem[{{Furuya} \& {Aikawa}(2018)}]{2018ApJ...857..105F}
{Furuya}, K. \& {Aikawa}, Y. 2018, \apj, 857, 105

\bibitem[{{Galli} {et~al.}(2018){Galli}, {Loinard}, {Ortiz-L{\'e}on},
  {Kounkel}, {Dzib}, {Mioduszewski}, {Rodr{\'\i}guez}, {Hartmann}, {Teixeira},
  {Torres}, {Rivera}, {Boden}, {Evans}, {Brice{\~n}o}, {Tobin}, \&
  {Heyer}}]{2018ApJ...859...33G}
{Galli}, P. A.~B., {Loinard}, L., {Ortiz-L{\'e}on}, G.~N., {et~al.} 2018, \apj,
  859, 33

\bibitem[{{Gerin} {et~al.}(2009){Gerin}, {Marcelino}, {Biver}, {Roueff},
  {Coudert}, {Elkeurti}, {Lis}, \& {Bockel{\'e}e-Morvan}}]{2009A&A...498L...9G}
{Gerin}, M., {Marcelino}, N., {Biver}, N., {et~al.} 2009, \aap, 498, L9

\bibitem[{{Griffin} {et~al.}(2010){Griffin}, {Abergel}, {Abreu}, {Ade},
  {Andr{\'e}}, {Augueres}, {Babbedge}, {Bae}, {Baillie}, {Baluteau}, {Barlow},
  {Bendo}, {Benielli}, {Bock}, {Bonhomme}, {Brisbin}, {Brockley-Blatt},
  {Caldwell}, {Cara}, {Castro-Rodriguez}, {Cerulli}, {Chanial}, {Chen},
  {Clark}, {Clements}, {Clerc}, {Coker}, {Communal}, {Conversi}, {Cox},
  {Crumb}, {Cunningham}, {Daly}, {Davis}, {de Antoni}, {Delderfield}, {Devin},
  {di Giorgio}, {Didschuns}, {Dohlen}, {Donati}, {Dowell}, {Dowell}, {Duband},
  {Dumaye}, {Emery}, {Ferlet}, {Ferrand}, {Fontignie}, {Fox}, {Franceschini},
  {Frerking}, {Fulton}, {Garcia}, {Gastaud}, {Gear}, {Glenn}, {Goizel},
  {Griffin}, {Grundy}, {Guest}, {Guillemet}, {Hargrave}, {Harwit}, {Hastings},
  {Hatziminaoglou}, {Herman}, {Hinde}, {Hristov}, {Huang}, {Imhof}, {Isaak},
  {Israelsson}, {Ivison}, {Jennings}, {Kiernan}, {King}, {Lange}, {Latter},
  {Laurent}, {Laurent}, {Leeks}, {Lellouch}, {Levenson}, {Li}, {Li},
  {Lilienthal}, {Lim}, {Liu}, {Lu}, {Madden}, {Mainetti}, {Marliani}, {McKay},
  {Mercier}, {Molinari}, {Morris}, {Moseley}, {Mulder}, {Mur}, {Naylor},
  {Nguyen}, {O'Halloran}, {Oliver}, {Olofsson}, {Olofsson}, {Orfei}, {Page},
  {Pain}, {Panuzzo}, {Papageorgiou}, {Parks}, {Parr-Burman}, {Pearce},
  {Pearson}, {P{\'e}rez-Fournon}, {Pinsard}, {Pisano}, {Podosek}, {Pohlen},
  {Polehampton}, {Pouliquen}, {Rigopoulou}, {Rizzo}, {Roseboom}, {Roussel},
  {Rowan-Robinson}, {Rownd}, {Saraceno}, {Sauvage}, {Savage}, {Savini},
  {Sawyer}, {Scharmberg}, {Schmitt}, {Schneider}, {Schulz}, {Schwartz},
  {Shafer}, {Shupe}, {Sibthorpe}, {Sidher}, {Smith}, {Smith}, {Smith},
  {Spencer}, {Stobie}, {Sudiwala}, {Sukhatme}, {Surace}, {Stevens}, {Swinyard},
  {Trichas}, {Tourette}, {Triou}, {Tseng}, {Tucker}, {Turner}, {Vaccari},
  {Valtchanov}, {Vigroux}, {Virique}, {Voellmer}, {Walker}, {Ward}, {Waskett},
  {Weilert}, {Wesson}, {White}, {Whitehouse}, {Wilson}, {Winter}, {Woodcraft},
  {Wright}, {Xu}, {Zavagno}, {Zemcov}, {Zhang}, \&
  {Zonca}}]{2010A&A...518L...3G}
{Griffin}, M.~J., {Abergel}, A., {Abreu}, A., {et~al.} 2010, \aap, 518, L3

\bibitem[{{Guzm{\'a}n} {et~al.}(2017){Guzm{\'a}n}, {{\"O}berg}, {Huang},
  {Loomis}, \& {Qi}}]{2017ApJ...836...30G}
{Guzm{\'a}n}, V.~V., {{\"O}berg}, K.~I., {Huang}, J., {Loomis}, R., \& {Qi}, C.
  2017, \apj, 836, 30

\bibitem[{{Hildebrand}(1983)}]{1983QJRAS..24..267H}
{Hildebrand}, R.~H. 1983, \qjras, 24, 267

\bibitem[{{Hily-Blant} {et~al.}(2013{\natexlab{a}}){Hily-Blant}, {Bonal},
  {Faure}, \& {Quirico}}]{2013Icar..223..582H}
{Hily-Blant}, P., {Bonal}, L., {Faure}, A., \& {Quirico}, E.
  2013{\natexlab{a}}, \icarus, 223, 582

\bibitem[{{Hily-Blant} {et~al.}(2019){Hily-Blant}, {Magalhaes de Souza},
  {Kastner}, \& {Forveille}}]{2019A&A...632L..12H}
{Hily-Blant}, P., {Magalhaes de Souza}, V., {Kastner}, J., \& {Forveille}, T.
  2019, \aap, 632, L12

\bibitem[{{Hily-Blant} {et~al.}(2020){Hily-Blant}, {Pineau des For{\^e}ts},
  {Faure}, \& {Flower}}]{2020A&A...643A..76H}
{Hily-Blant}, P., {Pineau des For{\^e}ts}, G., {Faure}, A., \& {Flower}, D.~R.
  2020, \aap, 643, A76

\bibitem[{{Hily-Blant} {et~al.}(2013{\natexlab{b}}){Hily-Blant}, {Pineau des
  For{\^e}ts}, {Faure}, {Le Gal}, \& {Padovani}}]{2013A&A...557A..65H}
{Hily-Blant}, P., {Pineau des For{\^e}ts}, G., {Faure}, A., {Le Gal}, R., \&
  {Padovani}, M. 2013{\natexlab{b}}, \aap, 557, A65

\bibitem[{Hunter(2007)}]{hunter2007}
Hunter, J.~D. 2007, Computing in Science \& Engineering, 9, 90

\bibitem[{{Ikeda} {et~al.}(2002){Ikeda}, {Hirota}, \&
  {Yamamoto}}]{2002ApJ...575..250I}
{Ikeda}, M., {Hirota}, T., \& {Yamamoto}, S. 2002, \apj, 575, 250

\bibitem[{{Jensen} {et~al.}(2021){Jensen}, {J{\o}rgensen}, {Kristensen},
  {Coutens}, {van Dishoeck}, {Furuya}, {Harsono}, \&
  {Persson}}]{2021A&A...650A.172J}
{Jensen}, S.~S., {J{\o}rgensen}, J.~K., {Kristensen}, L.~E., {et~al.} 2021,
  \aap, 650, A172

\bibitem[{{Jensen} {et~al.}(2023){Jensen}, {Spezzano}, {Caselli}, {Grassi}, \&
  {Haugb{\o}lle}}]{2023A&A...675A..34J}
{Jensen}, S.~S., {Spezzano}, S., {Caselli}, P., {Grassi}, T., \&
  {Haugb{\o}lle}, T. 2023, \aap, 675, A34

\bibitem[{{Juvela}(2020)}]{2020A&A...644A.151J}
{Juvela}, M. 2020, \aap, 644, A151

\bibitem[{{Loison} {et~al.}(2020){Loison}, {Wakelam}, {Gratier}, \&
  {Hickson}}]{2020MNRAS.498.4663L}
{Loison}, J.-C., {Wakelam}, V., {Gratier}, P., \& {Hickson}, K.~M. 2020,
  \mnras, 498, 4663

\bibitem[{{Magalh{\~a}es} {et~al.}(2018){Magalh{\~a}es}, {Hily-Blant}, {Faure},
  {Hernandez-Vera}, \& {Lique}}]{2018A&A...615A..52M}
{Magalh{\~a}es}, V.~S., {Hily-Blant}, P., {Faure}, A., {Hernandez-Vera}, M., \&
  {Lique}, F. 2018, \aap, 615, A52

\bibitem[{{Marty} {et~al.}(2011){Marty}, {Chaussidon}, {Wiens}, {Jurewicz}, \&
  {Burnett}}]{2011Sci...332.1533M}
{Marty}, B., {Chaussidon}, M., {Wiens}, R.~C., {Jurewicz}, A.~J.~G., \&
  {Burnett}, D.~S. 2011, Science, 332, 1533

\bibitem[{{Milam} {et~al.}(2005){Milam}, {Savage}, {Brewster}, {Ziurys}, \&
  {Wyckoff}}]{2005ApJ...634.1126M}
{Milam}, S.~N., {Savage}, C., {Brewster}, M.~A., {Ziurys}, L.~M., \& {Wyckoff},
  S. 2005, \apj, 634, 1126

\bibitem[{{M{\"u}ller} {et~al.}(2001){M{\"u}ller}, {Thorwirth}, {Roth}, \&
  {Winnewisser}}]{2001A&A...370L..49M}
{M{\"u}ller}, H.~S.~P., {Thorwirth}, S., {Roth}, D.~A., \& {Winnewisser}, G.
  2001, \aap, 370, L49

\bibitem[{{Mullins} {et~al.}(2016){Mullins}, {Loughnane}, {Redman}, {Wiles},
  {Guegan}, {Barrett}, \& {Keto}}]{2016MNRAS.459.2882M}
{Mullins}, A.~M., {Loughnane}, R.~M., {Redman}, M.~P., {et~al.} 2016, \mnras,
  459, 2882

\bibitem[{{Nier}(1950)}]{1950PhRv...77..789N}
{Nier}, A.~O. 1950, Physical Review, 77, 789

\bibitem[{{Priestley} {et~al.}(2023){Priestley}, {Whitworth}, \&
  {Fogerty}}]{2023MNRAS.518.4839P}
{Priestley}, F.~D., {Whitworth}, A.~P., \& {Fogerty}, E. 2023, \mnras, 518,
  4839

\bibitem[{{Redaelli} {et~al.}(2020){Redaelli}, {Bizzocchi}, \&
  {Caselli}}]{2020A&A...644A..29R}
{Redaelli}, E., {Bizzocchi}, L., \& {Caselli}, P. 2020, \aap, 644, A29

\bibitem[{{Redaelli} {et~al.}(2018){Redaelli}, {Bizzocchi}, {Caselli}, {Harju},
  {Chac{\'o}n-Tanarro}, {Leonardo}, \& {Dore}}]{2018A&A...617A...7R}
{Redaelli}, E., {Bizzocchi}, L., {Caselli}, P., {et~al.} 2018, \aap, 617, A7

\bibitem[{{Redaelli} {et~al.}(2023){Redaelli}, {Bizzocchi}, {Caselli}, \&
  {Pineda}}]{2023A&A...674L...8R}
{Redaelli}, E., {Bizzocchi}, L., {Caselli}, P., \& {Pineda}, J.~E. 2023, \aap,
  674, L8

\bibitem[{{Schnee} {et~al.}(2010){Schnee}, {Enoch}, {Noriega-Crespo}, {Sayers},
  {Terebey}, {Caselli}, {Foster}, {Goodman}, {Kauffmann}, {Padgett}, {Rebull},
  {Sargent}, \& {Shetty}}]{2010ApJ...708..127S}
{Schnee}, S., {Enoch}, M., {Noriega-Crespo}, A., {et~al.} 2010, \apj, 708, 127

\bibitem[{{Sch{\"o}ier} {et~al.}(2005){Sch{\"o}ier}, {van der Tak}, {van
  Dishoeck}, \& {Black}}]{2005A&A...432..369S}
{Sch{\"o}ier}, F.~L., {van der Tak}, F.~F.~S., {van Dishoeck}, E.~F., \&
  {Black}, J.~H. 2005, \aap, 432, 369

\bibitem[{{Shinnaka} {et~al.}(2014){Shinnaka}, {Kawakita}, {Kobayashi},
  {Nagashima}, \& {Boice}}]{2014ApJ...782L..16S}
{Shinnaka}, Y., {Kawakita}, H., {Kobayashi}, H., {Nagashima}, M., \& {Boice},
  D.~C. 2014, \apjl, 782, L16

\bibitem[{{Sipil{\"a}} {et~al.}(2023){Sipil{\"a}}, {Colzi}, {Roueff},
  {Caselli}, {Fontani}, \& {Wirstr{\"o}m}}]{2023A&A...678A.120S}
{Sipil{\"a}}, O., {Colzi}, L., {Roueff}, E., {et~al.} 2023, \aap, 678, A120

\bibitem[{{Spezzano} {et~al.}(2016){Spezzano}, {Bizzocchi}, {Caselli}, {Harju},
  \& {Br{\"u}nken}}]{2016A&A...592L..11S}
{Spezzano}, S., {Bizzocchi}, L., {Caselli}, P., {Harju}, J., \& {Br{\"u}nken},
  S. 2016, \aap, 592, L11

\bibitem[{{Spezzano} {et~al.}(2022){Spezzano}, {Caselli}, {Sipil{\"a}}, \&
  {Bizzocchi}}]{2022A&A...664L...2S}
{Spezzano}, S., {Caselli}, P., {Sipil{\"a}}, O., \& {Bizzocchi}, L. 2022, \aap,
  664, L2

\bibitem[{Virtanen {et~al.}(2020)Virtanen, Gommers, Oliphant, Haberland, Reddy,
  Cournapeau, Burovski, Peterson, Weckesser, Bright, {van der Walt}, Brett,
  Wilson, Millman, Mayorov, Nelson, Jones, Kern, Larson, Carey, Polat, Feng,
  Moore, {VanderPlas}, Laxalde, Perktold, Cimrman, Henriksen, Quintero, Harris,
  Archibald, Ribeiro, Pedregosa, {van Mulbregt}, \& {SciPy 1.0
  Contributors}}]{2020SciPy-NMeth}
Virtanen, P., Gommers, R., Oliphant, T.~E., {et~al.} 2020, Nature Methods, 17,
  261

\bibitem[{{Wampfler} {et~al.}(2014){Wampfler}, {J{\o}rgensen}, {Bizzarro}, \&
  {Bisschop}}]{2014A&A...572A..24W}
{Wampfler}, S.~F., {J{\o}rgensen}, J.~K., {Bizzarro}, M., \& {Bisschop}, S.~E.
  2014, \aap, 572, A24

\bibitem[{{Wirstr{\"o}m} \& {Charnley}(2018)}]{2018MNRAS.474.3720W}
{Wirstr{\"o}m}, E.~S. \& {Charnley}, S.~B. 2018, \mnras, 474, 3720

\bibitem[{{Yoshida} {et~al.}(2019){Yoshida}, {Sakai}, {Nishimura}, {Tokudome},
  {Watanabe}, {Sakai}, {Takano}, \& {Yamamoto}}]{2019PASJ...71S..18Y}
{Yoshida}, K., {Sakai}, N., {Nishimura}, Y., {et~al.} 2019, \pasj, 71, S18

\end{thebibliography}
\begin{appendix}

\section{Table of $^{14}$N/$^{15}$N ratios in HCN}\label{app:references}
Values reported in Figs. \ref{fig:Nfrac} and \ref{fig:Nfrac_NEW} are presented in Table \ref{table:nfrac}, along with the method used and references.
\begin{table*}
\caption{Measured HC$^{14}$N/HC$^{15}$N ratios for various sources.}             
\label{table:nfrac}      
\centering 
\smallskip \smallskip

\begin{tabular}{l c c c}        
\hline\hline                 
            \noalign{\smallskip} 

Object & $^{14}$N/$^{15}$N &  Method & Reference \\
\hline                        
            \noalign{\smallskip} 
  \multicolumn{4}{c}{\it Starless and pre-stellar cores} \\
              \noalign{\smallskip} 
              
\hline
CB23 & 216$\pm$60 & direct & This work \\
TMC2 & 461$\pm$130 & direct & This work \\
L1495 & 197$\pm$61 & direct & This work \\
L1512 & 251$\pm$76 & direct & This work \\
L1517B & 280$\pm$85 & direct & This work \\
L1495AN & 340$\pm$53 & indirect & This work \\
L183 & 195$\pm$55 & indirect & 1 \\
L1544 & 250$\pm$110 & indirect & 1 \\ 
B1-b & 165$\pm$30 & direct & 2 \\
L1498 & 338$\pm$28 & direct & 3 \\
L1521E & 150$\pm$50 & indirect & 4 \\
\hline                        
            \noalign{\smallskip} 
  \multicolumn{4}{c}{\it Protostars} \\
  \noalign{\smallskip} 
\hline
L483 & 321$\pm$96 & direct & 5 \\
L1527 & 250$\pm$80 & indirect & 6 \\
IRAS16293 & 200$\pm$40 & indirect & 7 \\
OMC-3 & 315$\pm$45 & indirect & 7 \\

\hline                        

            \noalign{\smallskip} 
  \multicolumn{4}{c}{\it Protoplanetary disks} \\
  \noalign{\smallskip} 
\hline
TW Hya & 223$\pm$21 & direct & 8 \\
HD163296 & 142$\pm$59 & indirect & 9 \\
MWC480 & 123$\pm$45 & indirect & 9 \\
V4046Sgr & 115$\pm$35 & indirect & 9 \\
LkCa15 & 83$\pm$32 & indirect & 9 \\
AS209 & 156$\pm$71 & indirect & 9 \\
\hline                        
            \noalign{\smallskip} 
  \multicolumn{4}{c}{\it Comets} \\
  \noalign{\smallskip} 
\hline
C/1995 O1 & 205$\pm$70 & indirect & 10 \\
17P/Holmes & 139$\pm$26 & indirect & 10 \\
        \noalign{\smallskip} 
        \hline                        
\end{tabular}
\tablefoot{For L1544, a more recent value is available based on maps presented in \cite{2022A&A...664L...2S}. These maps show a spatial difference in the fractionation ratio and we choose to include the single pointing value in this table for simplicity.}
\tablebib{
(1) \citet{2013Icar..223..582H} (2) \citet{2013A&A...560A...3D}; (3) \citet{2018A&A...615A..52M}; (4) \citet{2002ApJ...575..250I}; (5) \citet{2019A&A...625A.147A}; (6) \citet{2019PASJ...71S..18Y}; (7) \citet{2014A&A...572A..24W}; (8) \citet{2019A&A...632L..12H}; (9) \citet{2017ApJ...836...30G}; (10) \citet{2008ApJ...679L..49B}.
   }
\end{table*}

\section{Corner plots}\label{app:cornerplots}

\begin{figure*}[ht]
\resizebox{\hsize}{!}
        {\includegraphics{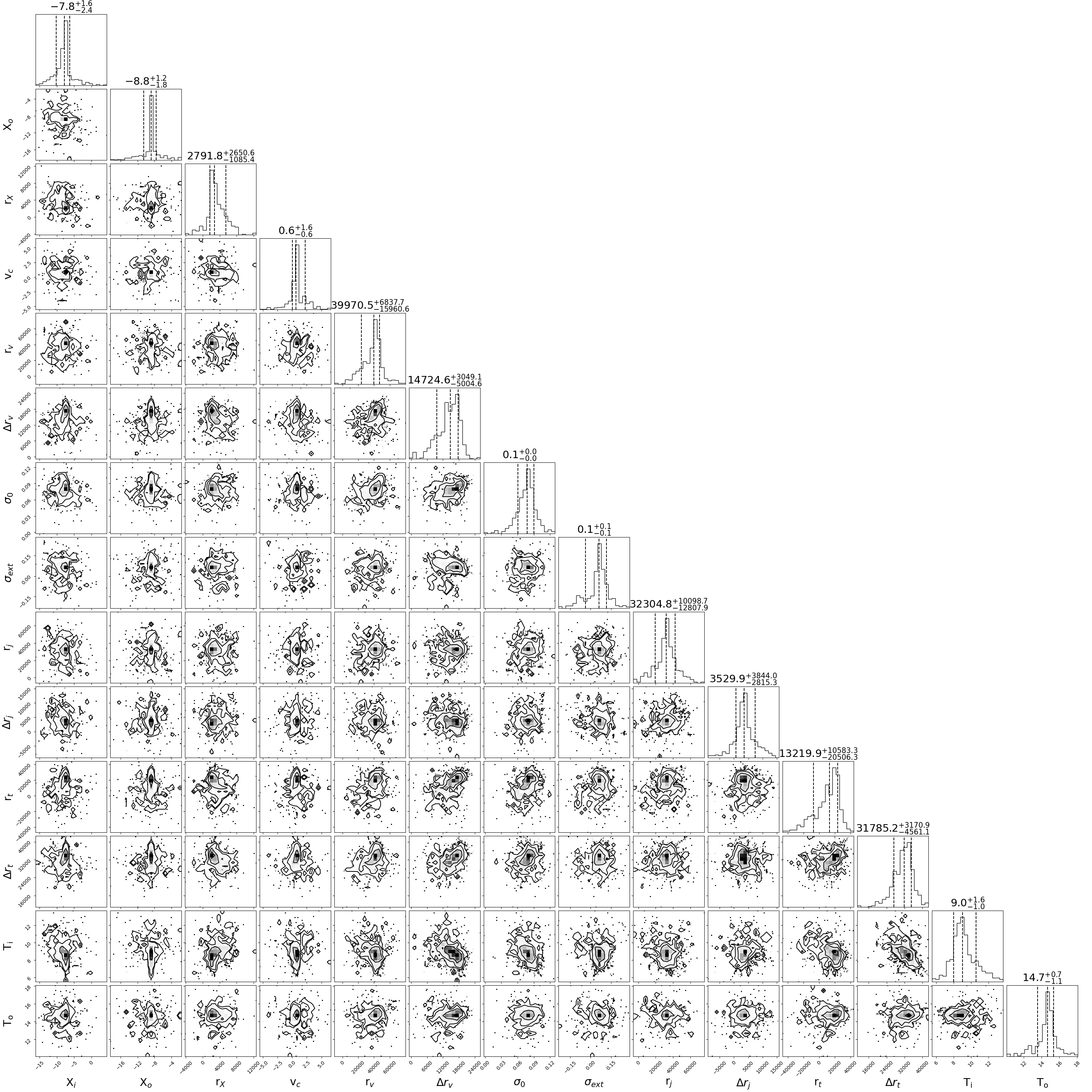}}
  \caption{Corner plot for the fitting of HCN and the physical structure of the core for CB23.}
     \label{fig:corner_CB23}
\end{figure*}

\begin{figure*}[ht]
\resizebox{\hsize}{!}
        {\includegraphics{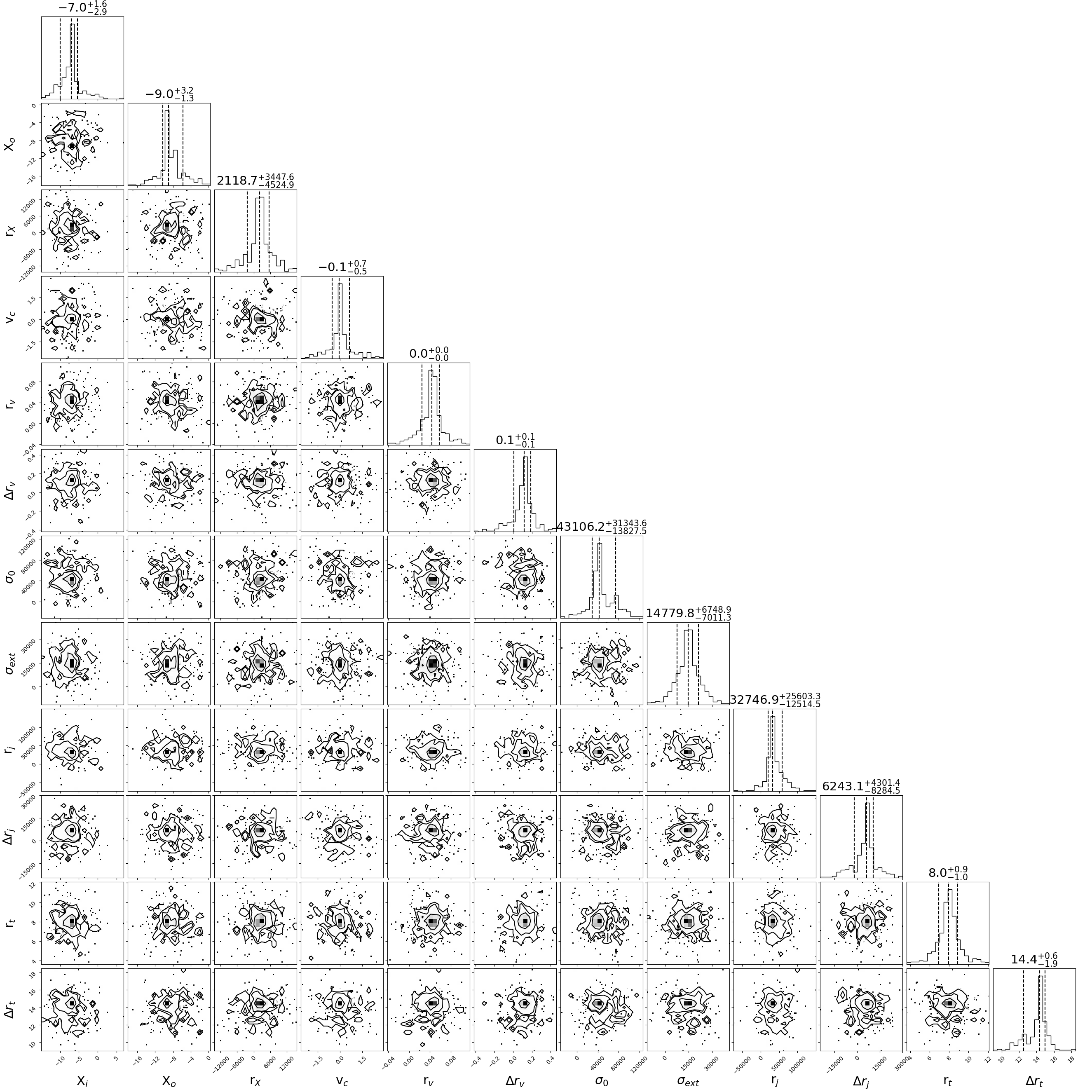}}
  \caption{Corner plot for the fitting of HCN and the physical structure of the core for L1495.}
     \label{fig:corner_L1495}
\end{figure*}

\begin{figure*}[ht]
\resizebox{\hsize}{!}
        {\includegraphics{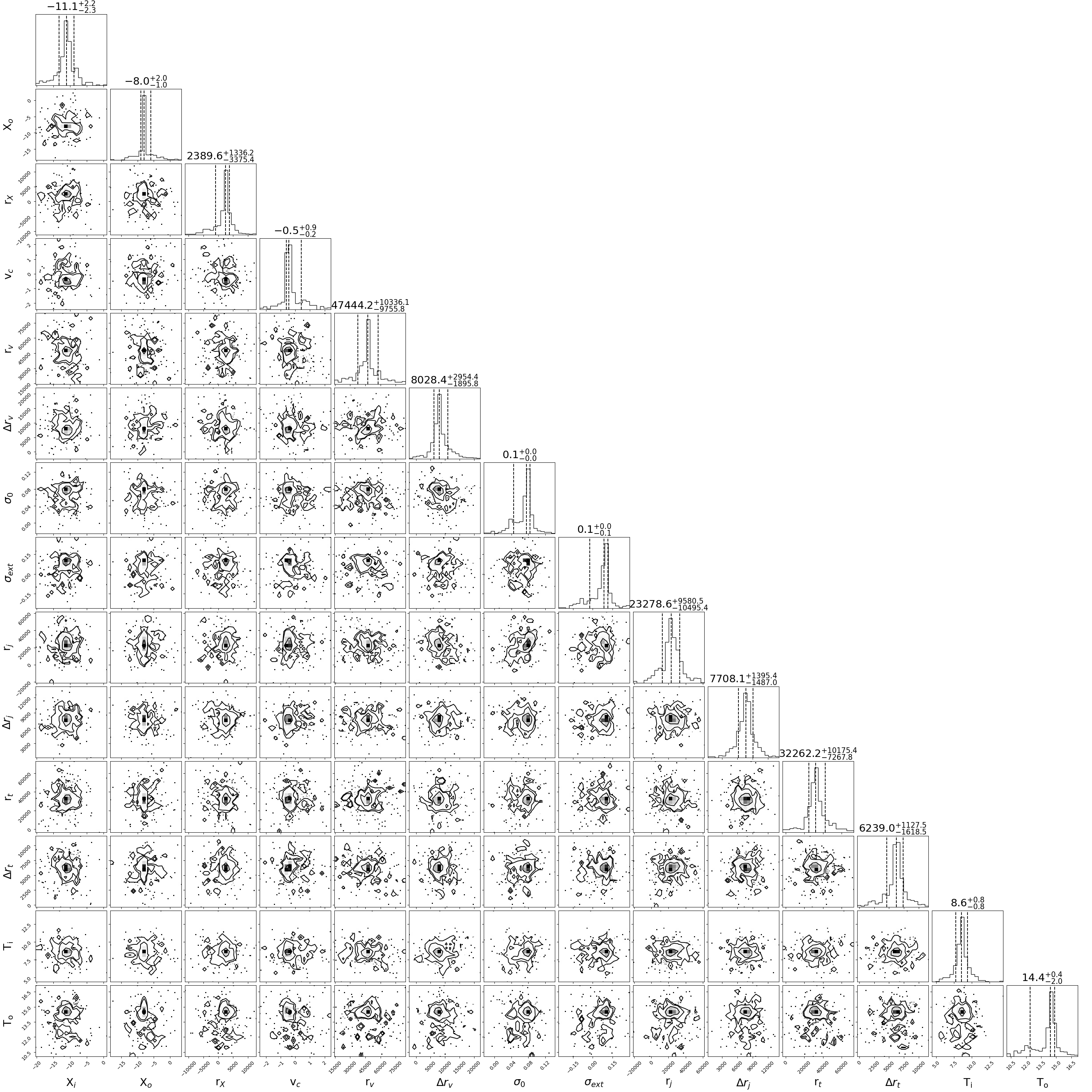}}
  \caption{Corner plot for the fitting of HCN and the physical structure of the core for L1512.}
     \label{fig:corner_L1512}
\end{figure*}

\begin{figure*}[ht]
\resizebox{\hsize}{!}
        {\includegraphics{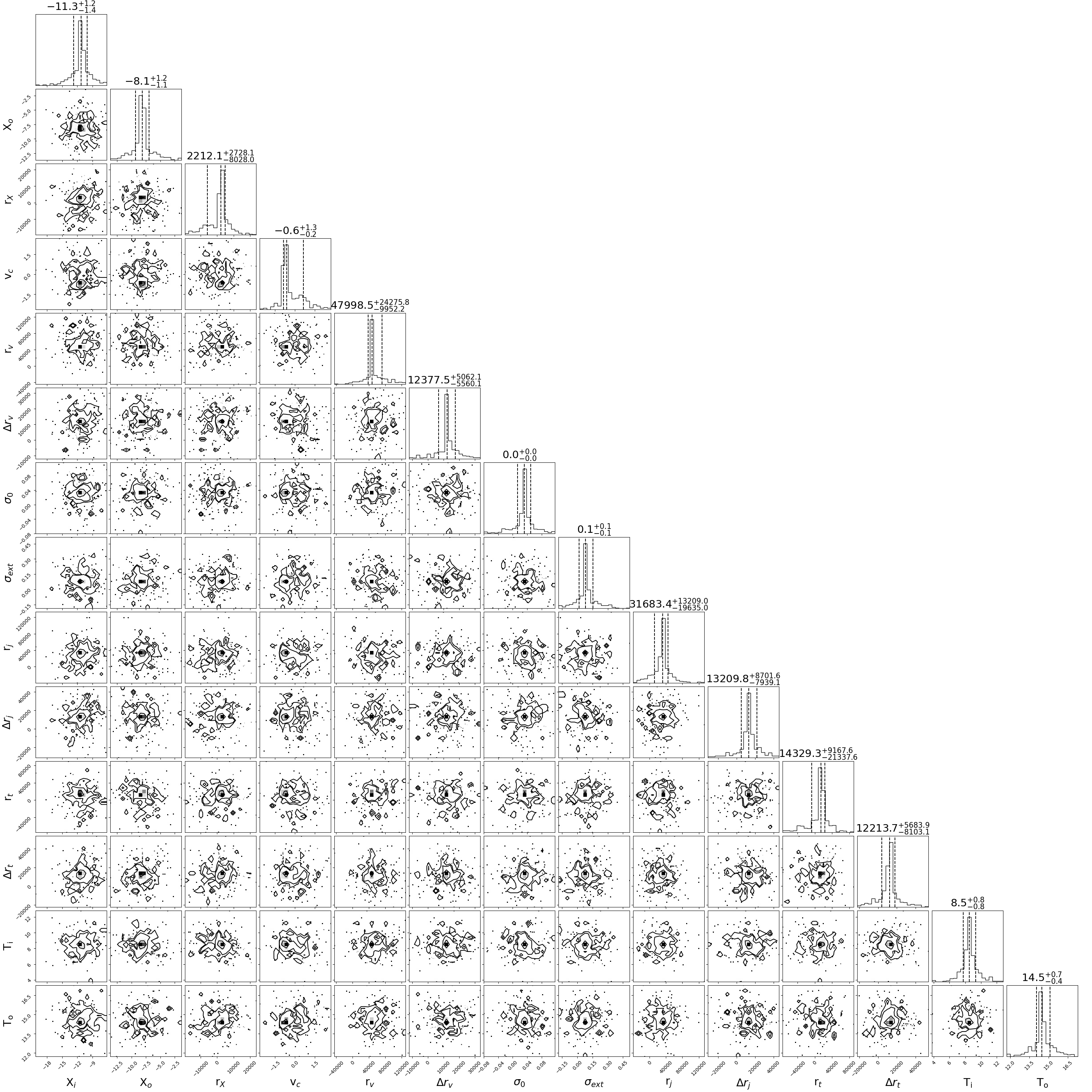}}
  \caption{Corner plot for the fitting of HCN and the physical structure of the core for L1517B.}
     \label{fig:corner_L1517B}
\end{figure*}

\end{appendix}
\end{document}